\documentclass[pre,twocolumn,amsmath,amssymb,superscriptaddress,showpacs]{revtex4}

\usepackage{graphicx}
\usepackage{latexsym}
\usepackage{amsmath}
\usepackage{amssymb}
\usepackage{amsfonts}
\usepackage{bm}

\advance\voffset 1.5cm

\newcommand{\la}{\left<}
\newcommand{\ra}{\right>}

\newcommand{\ddiff}{\ensuremath{\text{d}}}

\newcommand{\kB}{\mbox{$k_{\rm B}$}}
\newcommand{\kBT}{\mbox{$k_{\rm B}T$}}

\newcommand{\Tglass}{\mbox{$T_{\rm g}$}}

\newcommand{\Ahat}{\ensuremath{\hat{A}}}
\newcommand{\dAhat}{\ensuremath{\delta\hat{A}}}
\newcommand{\Bhat}{\ensuremath{\hat{B}}}
\newcommand{\dBhat}{\ensuremath{\delta\hat{B}}}
\newcommand{\Xhat}{\ensuremath{\hat{X}}}

\newcommand{\Vhat}{\ensuremath{\hat{V}}}
\newcommand{\dVhat}{\ensuremath{\delta\hat{V}}}
\newcommand{\Phat}{\ensuremath{\hat{P}}}
\newcommand{\dPhat}{\ensuremath{\delta\hat{P}}}
\newcommand{\Pid}{\ensuremath{P_\mathrm{id}}}
\newcommand{\Pidhat}{\ensuremath{\hat{P}_\mathrm{id}}}
\newcommand{\Pex}{\ensuremath{P_\mathrm{ex}}}
\newcommand{\Pexhat}{\ensuremath{\hat{P}_\mathrm{ex}}}

\newcommand{\Fex}{\ensuremath{F_\mathrm{ex}}}
\newcommand{\Zid}{\ensuremath{Z_\mathrm{id}}}
\newcommand{\Zex}{\ensuremath{Z_\mathrm{ex}}}
\newcommand{\Kid}{\ensuremath{K_\mathrm{id}}}
\newcommand{\Kex}{\ensuremath{K_\mathrm{ex}}}
\newcommand{\KdvolP}{\ensuremath{\left.{\cal F}_\mathrm{vol}\right|_0}}
\newcommand{\Krowl}{\ensuremath{{\cal F}_\mathrm{Row}}}
\newcommand{\KrowlV}{\ensuremath{\left.{\cal F}_\mathrm{Row}\right|_1}}
\newcommand{\KrowlP}{\ensuremath{\left.{\cal F}_\mathrm{Row}\right|_0}}
\newcommand{\KrowlL}{\ensuremath{\left.{\cal F}_\mathrm{Row}\right|_{\lambda}}}
\newcommand{\KmodVP}{\ensuremath{{\cal F}_\mathrm{reg}}}
\newcommand{\cmodVP}{\ensuremath{c_\mathrm{reg}}}
\newcommand{\cmodVPP}{\ensuremath{\left.c_\mathrm{reg}\right|_0}}
\newcommand{\cmodVPL}{\ensuremath{\left.c_\mathrm{reg}\right|_{\lambda}}}
\newcommand{\Hams}{\ensuremath{{\cal H}_s}}
\newcommand{\Uexs}{\ensuremath{U_s}}

\newcommand{\etaA}{\ensuremath{\eta_\mathrm{A}}}
\newcommand{\etaAid}{\ensuremath{\eta_\mathrm{A,id}}}
\newcommand{\etaAex}{\ensuremath{\eta_\mathrm{A,ex}}}
\newcommand{\etaAexhat}{\ensuremath{\hat{\eta}_\mathrm{A,ex}}}
\newcommand{\etaB}{\ensuremath{\eta_\mathrm{Born}}}

\newcommand{\etaBV}{\ensuremath{\left.\eta_\mathrm{Born}\right|_1}}
\newcommand{\etaBP}{\ensuremath{\left.\eta_\mathrm{Born}\right|_0}}
\newcommand{\etaF}{\ensuremath{\eta_\mathrm{F}}}
\newcommand{\etaFV}{\ensuremath{\left.\eta_\mathrm{F}\right|_1}}
\newcommand{\etaFP}{\ensuremath{\left.\eta_\mathrm{F}\right|_0}}
\newcommand{\etaFL}{\ensuremath{\left.\eta_\mathrm{F}\right|_{\lambda}}}
\newcommand{\etaFid}{\ensuremath{\eta_\mathrm{F,id}}}

\newcommand{\etaFidP}{\ensuremath{\left.\eta_\mathrm{F,id}\right|_0}}
\newcommand{\etaFidL}{\ensuremath{\left.\eta_\mathrm{F,id}\right|_{\lambda}}}
\newcommand{\etaFex}{\ensuremath{\eta_\mathrm{F,ex}}}
\newcommand{\etaFexV}{\ensuremath{\left.\eta_\mathrm{F,ex}\right|_1}}
\newcommand{\etaFexP}{\ensuremath{\left.\eta_\mathrm{F,ex}\right|_0}}
\newcommand{\etaFexL}{\ensuremath{\left.\eta_\mathrm{F,ex}\right|_{\lambda}}}
\newcommand{\etaFmix}{\ensuremath{\eta_\mathrm{F,mix}}}

\newcommand{\etaFmixP}{\ensuremath{\left.\eta_\mathrm{F,mix}\right|_0}}
\newcommand{\etaFmixL}{\ensuremath{\left.\eta_\mathrm{F,mix}\right|_{\lambda}}}

\newcommand{\NVT}{\ensuremath{\text{NVT}}}
\newcommand{\NPT}{\ensuremath{\text{NPT}}}

\newcommand{\dk}{\ensuremath{\delta k}}
\newcommand{\kl}{\ensuremath{k_l}}
\newcommand{\Rl}{\ensuremath{R_l}}
\newcommand{\rl}{\ensuremath{r_l}}
\newcommand{\xl}{\ensuremath{x_l}}
\newcommand{\dVtyp}{\ensuremath{\delta V_0}}
\newcommand{\dVtypsq}{\ensuremath{\delta V^2_0}}
\newcommand{\fl}{\ensuremath{f_l}}

\newcommand{\nla}{\ensuremath{n_{l}^{\alpha}}}
\newcommand{\nlb}{\ensuremath{n_{l}^{\beta}}}
\newcommand{\nlc}{\ensuremath{n_{l}^{\gamma}}}
\newcommand{\nld}{\ensuremath{n_{l}^{\delta}}}
\newcommand{\na}{\ensuremath{n^{\alpha}}}
\newcommand{\nb}{\ensuremath{n^{\beta}}}
\newcommand{\nc}{\ensuremath{n^{\gamma}}}
\newcommand{\nd}{\ensuremath{n^{\delta}}}

\newcommand{\Aab}{\ensuremath{A^{\alpha\beta}}}
\newcommand{\Aabcd}{\ensuremath{A^{\alpha\beta\gamma\delta}}}

\newcommand{\Babcd}{\ensuremath{B^{\alpha\beta\gamma\delta}}}

\newcommand{\Cidabcd}{\ensuremath{C_\mathrm{id}^{\alpha\beta\gamma\delta}}}
\newcommand{\CKabcd}{\ensuremath{C_\mathrm{K}^{\alpha\beta\gamma\delta}}}
\newcommand{\CBabcd}{\ensuremath{C_\mathrm{Born}^{\alpha\beta\gamma\delta}}}
\newcommand{\CFabcd}{\ensuremath{C_\mathrm{F}^{\alpha\beta\gamma\delta}}}
\newcommand{\CFidabcd}{\ensuremath{C_\mathrm{F,id}^{\alpha\beta\gamma\delta}}}
\newcommand{\CFexabcd}{\ensuremath{C_\mathrm{F,ex}^{\alpha\beta\gamma\delta}}}
\newcommand{\Eabcd}{\ensuremath{E^{\alpha\beta\gamma\delta}}}

\newcommand{\Ttwo}{{\bf \ensuremath{{\cal T}_2}}}
\newcommand{\Tfour}{{\bf \ensuremath{{\cal T}_4}}}

\newcommand{\sigmaTen}{\ensuremath{\sigma^{\alpha\beta}}}
\newcommand{\sigmaidTen}{\ensuremath{\sigma_\mathrm{id}^{\alpha\beta}}}
\newcommand{\sigmaexTen}{\ensuremath{\sigma_\mathrm{ex}^{\alpha\beta}}}
\newcommand{\epsilonTen}{\ensuremath{\epsilon^{\alpha\beta}}}

\newcommand{\pressexTen}{\ensuremath{\hat{P}_\mathrm{ex}^{\alpha\beta}}}

\newcommand{\fscalP}{\ensuremath{f_0}}

\newcommand{\fext}{\ensuremath{f_\text{ext}}}
\newcommand{\kext}{\ensuremath{k_\text{ext}}}
\newcommand{\Kext}{\ensuremath{K_\text{ext}}}
\newcommand{\Uext}{\ensuremath{U_\text{ext}}}
\newcommand{\Vext}{\ensuremath{V_\text{ext}}}

\newcommand{\rhoref}{\ensuremath{\rho_\text{ref}}}
\newcommand{\Kref}{\ensuremath{K_\text{ref}}}
\newcommand{\Vref}{\ensuremath{V_\text{ref}}}

\newcommand{\plam}{\ensuremath{p}}

\newcommand{\dVtyplam}{\ensuremath{\delta V^2_{\lambda}}}

\begin{document}

\title{Compressibility and pressure correlations in isotropic solids and fluids}

\author{J.P.~Wittmer}
\email{joachim.wittmer@ics-cnrs.unistra.fr}
\affiliation{Institut Charles Sadron, Universit\'e de Strasbourg \& CNRS, 23 rue du Loess, 67034 Strasbourg Cedex, France}
\author{H.~Xu}
\affiliation{LCP-A2MC, Institut Jean Barriol, Universit\'e de Lorraine \& CNRS,\\ 1 bd Arago, 57078 Metz Cedex 03, France}
\author{P.~Poli\'nska}
\affiliation{Institut Charles Sadron, Universit\'e de Strasbourg \& CNRS, 23 rue du Loess, 67034 Strasbourg Cedex, France}
\author{C. Gillig}
\affiliation{FMF, University of Freiburg, Stefan-Meier-Str. 21, D-79104 Freiburg, Germany}
\author{J. Helfferich}
\affiliation{Theoretical Polymer Physics, University of Freiburg, Hermann-Herder-Str. 3, D-79104 Freiburg, Germany}
\author{F.~Weysser}
\affiliation{Institut Charles Sadron, Universit\'e de Strasbourg \& CNRS, 23 rue du Loess, 67034 Strasbourg Cedex, France}
\author{J. Baschnagel}
\affiliation{Institut Charles Sadron, Universit\'e de Strasbourg \& CNRS, 23 rue du Loess, 67034 Strasbourg Cedex, France}

\begin{abstract}
Presenting simple coarse-grained models of isotropic solids and fluids in $d=1$, $2$ and $3$ dimensions 
we investigate the correlations of the instantaneous pressure and its ideal and excess contributions
at either imposed pressure (\NPT-ensemble, $\lambda=0$) or volume (\NVT-ensemble, $\lambda=1$)
and for more general values of the dimensionless parameter $\lambda$ characterizing the constant-volume constraint.
The stress fluctuation representation $\left.\Krowl\right|_{\lambda=1}$ of the compression 
modulus $K$ in the \NVT-ensemble is derived directly (without a microscopic displacement field) 
using the well-known thermodynamic transformation rules between conjugated ensembles. 
The transform is made manifest by computing the Rowlinson functional $\Krowl$ also in the \NPT-ensemble 
where $\left.\Krowl\right|_{\lambda=0} = K \fscalP(x)$ with $x = \Pid/K$ being a scaling variable, 
$\Pid$ the ideal pressure and $\fscalP(x) = x (2-x)$ a universal function. 
By gradually increasing $\lambda$ by means of an external spring potential, the crossover between both 
classical ensemble limits is monitored. 
This demonstrates, e.g., the lever rule $\KrowlL = K \left[ \lambda + (1-\lambda) \fscalP(x) \right]$.
\end{abstract}


\date{\today}
\maketitle



\section{Introduction}
\label{sec_intro}

\paragraph*{Strain and stress fluctuations.}
Among the fundamental properties of any equilibrium system are its (generalized) elastic 
constants characterizing the fluctuations of its extensive and/or conjugated intensive variables 
\cite{LandauElasticity,Callen,RowlinsonBook,Hoover69,Lutsko89,WTBL02,Pablo03,Barrat88,Barrat08}.
For instance, for an isotropic solid or fluid the volume and density fluctuations are set by the 
isothermal compression modulus $K$ defined as \cite{Callen}
\begin{equation}
K \equiv V \left.\frac{\partial^2 F}{\partial V^2}\right|_T \equiv 
- V \left.\frac{\partial P}{\partial V}\right|_T = 
\left.\rho \frac{\partial P}{\partial \rho}\right|_T
\label{eq_Kdef}
\end{equation}
with $V$ being the volume, $N$ the particle number, $\rho = N/V$ the particle density,
$F(T,V)$ the free energy, $P$ the (mean) pressure and $T$ the (mean) temperature. Albeit $K$ may in principle 
be measured by fitting the pressure isotherm $P(\rho,T)$ \cite{SXM12},
it is from the computational point of view important \cite{AllenTildesleyBook,FrenkelSmitBook}
that this modulus may be obtained from the volume fluctuations at constant pressure ($\NPT$-ensemble)
and the pressure fluctuations at constant volume ($\NVT$-ensemble) 
evaluated at the {\em same} state point, i.e. at the same mean temperature, density and pressure \cite{Callen}.
%
In the \NPT-ensemble $K$ is obtained from the fluctuations  $\delta \Vhat = \Vhat - V$ 
of the instantaneous volume $\Vhat$ around its 
mean value $V = \langle \Vhat \rangle$ using \cite{Callen}                   
\begin{equation}                                                                                       
K = \KdvolP \equiv                                                                                     
\kBT V / \left.\langle \delta \Vhat^2 \rangle\right|_0                                                
\label{eq_K_NPT}
\end{equation}
with $\kB$ being Boltzmann's constant 
and where $|_0$ indicates that the average is obtained in the \NPT-ensemble 
($\lambda = 0$ with $\lambda$ defined below).
%
%
Equivalently, $K$ may be obtained in a canonical $\NVT$-ensemble using the stress fluctuation formula 
\cite{RowlinsonBook,Hoover69,Lutsko89,AllenTildesleyBook,SXM12,WXP13}
\begin{equation}
K = \KrowlV \equiv P + \etaBV - \etaFexV
\label{eq_Rowl}
\end{equation}
with $|_1$ indicating the \NVT-ensemble ($\lambda = 1$) and other definitions given immediately below. 
This formula was first stated for liquids in the 1950s 
by Rowlinson \cite{RowlinsonBook} and later {\em implicitly} rediscovered 
by Squire, Hold and Hoover \cite{Hoover69} 
formulating the stress fluctuation formalism for anisotropic solids 
as summarized in appendix~\ref{app_generalmodul}.

\paragraph*{Affine contribution.}
The second term $\etaB$ in eq.~(\ref{eq_Rowl}) represents the Born approximation \cite{BornHuang}
for the interaction energy implied by an imposed infinitesimal strain 
assuming {\em affine} microscopic particle displacements \cite{Lutsko89,WTBL02,Barrat08}. 
As reminded in appendix~\ref{app_affineenergy}, 
for pairwise additive potentials this ``Born-Lam\'e coefficient" becomes \cite{AllenTildesleyBook}
\begin{equation}
\etaB \equiv 
\frac{1}{d^2 V} \la \sum_l h(\rl) \ra  
\mbox{ with } h(r) \equiv r \ (r \ u^{\prime}(r))^{\prime}
\label{eq_etaB}
\end{equation}
and $d$ being the spatial dimension, $l$ an index labeling the interactions, 
$\rl$ the distance between two interacting particles, $u(r)$ the pair potential
and a prime denoting a derivative with respect to the indicated variable. 
As suggested by eq.~(\ref{eq_etaB}), $\etaB$ is sometimes also called ``hypervirial" \cite{AllenTildesleyBook}.
We note {\em en passant} that $\etaB$ is a moment of the {\em second} derivative of $u(r)$ and some care is 
required if $\etaB$ is computed using a truncated potential \cite{XWP12,foot_trunc}.
%

\paragraph*{Non-affine contribution.}
In general the Born approximation overpredicts the free-energy change. 
The overprediction is ``corrected" by the stress fluctuation term
\begin{equation}
\etaFexV \equiv \beta V \left.\la \delta \Pexhat^2 \ra\right|_1 \ge 0
\label{eq_etaF}
\end{equation}
with $\beta = 1/\kBT$ being the inverse temperature and $\Pexhat$ the instantaneous 
excess pressure which for pairwise additive potentials is given by Kirkwood's virial
\cite{AllenTildesleyBook}
\begin{equation}
\Pexhat \equiv \frac{1}{d \Vhat} \sum_l \rl \fl 
\ \mbox{ with } \ \fl = - u^{\prime}(\rl)
\label{eq_PexKirkwood}
\end{equation}
being the central force between two interacting particles.
(Although $\Pexhat$ is used here in Rowlinson's formula, eq.~(\ref{eq_Rowl}), 
at constant volume $V = \Vhat$, we have written it in a slightly more general form 
which is necessary if volume fluctuations are allowed.)
In particular from ref.~\cite{Lutsko89} it has become clear that stress fluctuation corrections,
such as $\etaFexV$, do not necessarily vanish for $T \to 0$. 
As we shall also illustrate in the present paper,
this is due to the fact that the particle displacements need not 
follow an imposed macroscopic strain affinely \cite{WTBL02,TWLB02,TLWB04,LBTWB05,Barrat06,ZT13}. 
%
%
How important the non-affine motions are, depends on the system under consideration \cite{Barrat06}. 
While the elastic properties of crystals with one atom per unit cell are given by the Born term only, 
stress fluctuations are significant for crystals with more complex unit cells \cite{Lutsko89}. 
They become pronounced for polymer-like soft materials \cite{SXM12} and amorphous solids 
\cite{Lemaitre04,Barrat06,Barrat88,WTBL02,TWLB02,TLWB04,LBTWB05,Pablo04,SBM11,LTWB06,ZT13}.
%

\paragraph*{Fluctuations in conjugated ensembles.}
%
Focusing on the compression modulus $K$ we emphasize in this report that the numerically
more convenient stress fluctuation formalism may be obtained directly 
using the well-known thermodynamic transformation rules between conjugated ensembles 
\cite{Lebowitz67,foot_classical}. 
This point is crucial if the formalism is used in situations 
where no meaningful microscopic displacement field can be defined \cite{WXP13,foot_critical}. 
Computing Rowlinson's $\Krowl$  for \NPT-ensembles the general transform behind the formalism
can be made manifest. Elaborating a short comment \cite{errat},
we show that
\begin{eqnarray}
\KrowlP & \equiv & P + \etaBP - \etaFexP \nonumber \\
& = & K \ \fscalP(x)
\ \mbox{ with } \fscalP(x)  = x (2 - x)
\label{eq_key}
\end{eqnarray}
being the universal scaling function, $x \equiv \Pid/K$ the scaling variable and 
$\Pid$ the ideal pressure contribution. 

\paragraph*{Generalized $\lambda$-ensembles.}
%
It is straightforward to interpolate between
the \NPT- and the \NVT-ensemble by imposing an external spring potential
\begin{equation}
\Uext(\Vhat) = \frac{\kext}{2} \left(\Vhat - \Vext \right)^2 \mbox{ with } \Kext \equiv V \kext
\label{eq_Uext}
\end{equation}
being the associated compression modulus introduced for convenience \cite{foot_kextdim}.
(Our approach is conceptually similar to the so-called ``Gaussian ensemble" proposed some years ago by
Hetherington and others \cite{Hetherington87,CETT06} generalizing the Boltzmann weight of the canonical
ensemble by an exponential factor $\Uext(\hat{E}) \propto \hat{E}^2$ of the instantaneous energy $\hat{E}$.)
Throughout this work it is assumed that $\Kext \ge 0$, i.e. $\Uext(\Vhat)$ {\em reduces} the
volume fluctuations \cite{foot_Workum}.
Chosing the reference volume $\Vext$ equal to the average volume $V$ of the isobaric system 
at imposed $P$ allows, for symmetry reasons, to work at constant mean pressure irrespective 
of the strength of the external potential \cite{foot_generalensemble}.
The volume fluctuations may then be characterized using the dimensionless parameter 
$\lambda \equiv \Kext /(K+ \Kext)$.
(Since $\Kext \ge 0$, we have $0 \le \lambda \le 1$ in the current study.)
\NPT-ensemble statistics is expected if the external potential does not constrain the volume
fluctuations, i.e. $\lambda \to 0$, while \NVT-statistics should become
relevant in the opposite limit for $\Kext \to \infty$ and $\lambda \to 1$.
We shall monitor various properties, such as the Rowlinson formula $\KrowlL$, 
as a function of $\lambda$ and $x=\Pid/K$.
We demonstrate, e.g., the simple lever rule
\begin{equation}
\KrowlL = K \left[ \lambda + (1-\lambda) \fscalP(x)  \right]
\label{eq_KrowlL}
\end{equation}
which generalizes eq.~(\ref{eq_Rowl}) and eq.~(\ref{eq_key}) to arbitrary $\lambda$.

%
\paragraph*{Outline.}
In sect.~\ref{sec_theo} we consider theoretically various correlation functions of 
normal stress contributions in different $\lambda$-ensembles. 
We begin by summarizing in sect.~\ref{theo_dAdBtrans} the transformation relations for fluctuations 
between \NVT- and \NPT-ensembles. Equation~(\ref{eq_key}) is derived in sect.~\ref{theo_NPT}.
We turn then in sect.~\ref{theo_dAdBL} to the transformation relations for general $\lambda$ and 
demonstrate eq.~(\ref{eq_KrowlL}) in sect.~\ref{theo_corrL}. 
Our Monte Carlo (MC) and molecular dynamics (MD) simulations of several simple coarse-grained models 
in $d=1$, $2$ and $3$ dimensions are described in sect.~\ref{sec_algo}. 
Our theoretical predictions are then checked numerically in sect.~\ref{sec_simu}.
Several well-known but scattered theoretical statements are gathered in the appendix.

\section{Theoretical considerations}
\label{sec_theo}

\subsection{Fluctuations in \NVT- and \NPT-ensembles}
\label{theo_dAdBtrans}
As discussed in the literature \cite{Callen,AllenTildesleyBook,Lebowitz67}, a simple average $A = \langle \hat{A} \rangle$ 
of an observable ${\cal A}$ does not depend on the chosen ensemble, at least not if the system is large enough.
(We do thus not indicate normally in which ensemble the average has been taken.)
However, a correlation function $\langle \dAhat \dBhat \rangle$ of two observables ${\cal A}$ and ${\cal B}$ 
depends on whether $V$ or $P$ are imposed. As shown first by Lebowitz {\em et al.} in 1967 \cite{Lebowitz67}, 
one verifies that \cite{foot_classical,foot_intensive}
\begin{equation}
\left. \la \dAhat \dBhat \ra\right|_{1} = 
\left. \la \dAhat \dBhat \ra\right|_{0} - \frac{K}{\beta V} \ \frac{\partial A}{\partial P} \frac{\partial B}{\partial P}
\label{eq_dAdB}
\end{equation}
to leading order.
%
We note that the left hand-side of eq.~(\ref{eq_dAdB}) must vanish if at least one of the observables is
a function of $\Vhat$. In this case we have
\begin{equation}
\left. \la \dAhat \dBhat \ra\right|_{0} = \frac{K}{\beta V} \ \frac{\partial A}{\partial P} \frac{\partial B}{\partial P}.
\label{eq_dAdBnoV}
\end{equation}
One verifies for $\Ahat = \Bhat = \Vhat$ that eq.~(\ref{eq_dAdBnoV}) is consistent with eq.~(\ref{eq_K_NPT}).
For $\Ahat = \Vhat$ and $\Bhat = \Phat$ one obtains immediately the well-known relation \cite{Callen}
\begin{equation}
-\beta \left.\la \delta \Vhat \delta \Phat \ra\right|_0 = 1.
\label{eq_VPcorr}
\end{equation}
%
%
Similarly, one obtains for $\Ahat = \Bhat = \Vhat^n$ that 
\begin{equation}
\left.\la \delta (\Vhat^n)^2 \ra\right|_0
= \frac{K}{\beta V} \ \left(\frac{\partial \langle \Vhat^n\rangle}{\partial P}\right)^2
\approx \frac{1}{\beta K} n^2 V^{2n-1}
\label{eq_VinvfluctuA}
\end{equation}
where the steepest-descent approximation 
\begin{equation}
\la \Vhat^n \ra^{1/n} \approx \la \Vhat \ra = V
\label{eq_Vhatn}
\end{equation}
for simple averages has been made and $V/K = -\partial V / \partial P$ is used again. 
For the fluctuations of the inverse volume $1/\Vhat$ the latter result ($n=-1$) may be rewritten 
compactly using eq.~(\ref{eq_K_NPT}) as
\begin{equation}
V^4 \frac{\left.\la \delta (1/\Vhat)^2 \ra\right|_0}{\left.\la \delta \Vhat^2 \ra\right|_0} = 1
\label{eq_VinvfluctuAA}
\end{equation}
where we have changed $\approx$ to the equal sign for large systems.
That eq.~(\ref{eq_VinvfluctuA}) and eq.~(\ref{eq_VinvfluctuAA}) become exact for $V \to \infty$ 
can be also seen by using that the distribution of $\Vhat$ in the \NPT-ensemble is Gaussian.
With $\Ahat =  \Vhat^n$ and $\Bhat = \Phat$ one gets similarly
\begin{equation}
\left.\la \delta (\Vhat^n) \delta \Phat \ra\right|_0
= \frac{K}{\beta V} \ \frac{\partial \langle \Vhat^n\rangle}{\partial P}
\approx -n V^{n-1} \kBT 
\label{eq_VinvfluctuB}
\end{equation}
to leading order for $V \to \infty$ using the same approximation as above.
With eq.~(\ref{eq_VPcorr}) this gives for $n=-1$ the convenient cumulant
\begin{equation}
-V^2 \frac{\left.\la \delta (1/\Vhat) \delta \Phat \ra\right|_0}{\left.\la \delta \Vhat \delta \Phat \ra\right|_0} = 1.
\label{eq_VinvfluctuBB}
\end{equation}
%
The cumulants eqs.~(\ref{eq_VinvfluctuAA},\ref{eq_VinvfluctuBB}) 
have been used in the computational part of our work to check the precision of the barostat and 
to verify whether our configurations are sufficiently large for the investigated state point.
\subsection{Transformation of pressure auto-correlations}
\label{theo_etaFtrans}
Returning to eq.~(\ref{eq_dAdB}), this implies for $\Ahat=\Bhat=\Phat$ the transformation 
of the pressure fluctuations 
\begin{equation}
\beta V \left.\la \delta \Phat^2 \ra\right|_1 = \beta V \left.\la \delta \Phat^2 \ra\right|_0 - K,
\label{eq_Pfluctutrans}
\end{equation}
i.e. $K$ may be obtained by measuring the pressure fluctuations in both ensembles. 
As we shall show in paragraph~\ref{theo_Krowl}, the numerically more convenient Rowlinson expression 
$\KrowlV$ for $K$ can be derived directly from eq.~(\ref{eq_Pfluctutrans}) \cite{WXP13}.
In the following we use the more concise notation $\etaF \equiv \beta V \langle \delta \Phat^2 \rangle$
for the pressure fluctuations. $\etaFP$ is also called the ``affine dilatational elasticity" $\etaA$
\cite{WXP13} for reasons which will become obvious in sect.~\ref{theo_Krowl}. 
(See also appendix~\ref{app_affineenergy}.) 
Since $K > 0$ for a stable system \cite{Callen}, eq.~(\ref{eq_Pfluctutrans}) implies $\etaA \equiv \etaFP > \etaFV$.
Depending on the disorder, $\etaFV$ is, however, {\em not} a negligible contribution. 
%
%
To see this let us remind that $K$ can also be determined 
from the $(\Vhat,\Phat)$-data measured in an \NPT-ensemble using the linear regression relation
\begin{equation}
K = \left.\KmodVP\right|_0  \equiv \left.-V \la \dVhat \dPhat \ra\right|_{0} / \left.\la \dVhat^2 \ra\right|_{0}.
\label{eq_KmodVP}
\end{equation}
Please note that using eq.~(\ref{eq_VPcorr}) this reduces to eq.~(\ref{eq_K_NPT}). 
Associated to $\left.\KmodVP\right|_0$ is the dimensionless regression coefficient \cite{abramowitz}
\begin{equation}
\left.\cmodVP\right|_0  \equiv  \left. - \la \dVhat \dPhat \ra\right|_{0} / 
\sqrt{\left.\la \dVhat^2 \ra\right|_{0} \left.\la \dPhat^2 \ra\right|_{0}}
\label{eq_cmodVP}
\end{equation}
which can be also further simplified using eq.~(\ref{eq_VPcorr}).
Interestingly, using eq.~(\ref{eq_K_NPT}) and eq.~(\ref{eq_Pfluctutrans}) one sees that 
\begin{equation}
\left.\cmodVP\right|_0 = \sqrt{K/\etaA} = \sqrt{1 - \etaFV/\etaA},
\label{eq_cmodVP_etaA}
\end{equation}
i.e. the regression coefficient obtained at constant pressure determines the pressure fluctuations
at constant volume.
Only if the measured $(\Vhat,\Phat)$ are perfectly correlated, i.e. $\left.\cmodVP\right|_0 =1$, this implies $K = \etaA$
and $\etaFV =0$. In fact, for all non-trivial systems one always has
\begin{equation}
\left.\cmodVP\right|_0 < 1, \mbox{ thus } K < \etaA \mbox{ and } \etaFV > 0,
\label{eq_bound}
\end{equation}
i.e. the affine dilatational elasticity $\etaA$ sets only an {\em upper bound} to the compression modulus
and a theory which only contains the affine response must overpredict $K$.

\subsection{Rowlinson's formula rederived}
\label{theo_Krowl}

\paragraph*{MC-gauge.}
There is a considerable freedom for defining the instantaneous value of the pressure
$\Phat = \Pidhat + \Pexhat$ as long as its average $P = \Pid + \Pex$ does not change \cite{AllenTildesleyBook}.
It is convenient for the subsequent derivations and for our MC simulations 
(and not in conflict with the also presented MD simulations) 
to define the instantaneous ideal pressure $\Pidhat$ by \cite{AllenTildesleyBook}
\begin{equation}
\Pidhat = \kBT N /\Vhat \ \ \ \mbox{(MC-gauge).}
\label{eq_MCgauge}
\end{equation}
Within this ``MC-gauge" the thermal momentum fluctuations are assumed to be integrated out.
This leads to the usual prefactor 
\begin{equation}
\Vhat^{N} = \exp\left[ -\beta (-\kBT N \log(\Vhat)) \right]
\label{eq_VNfactor}
\end{equation}
of the remaining partition function.
The effective Hamiltonian $\Hams(\Vhat)$ of a state $s$ of the system thus reads
\begin{equation}
\Hams(\Vhat) = - \kBT N \log(\Vhat) + \Uexs(\Vhat) + \text{const}
\label{eq_Hams}
\end{equation}
where the first term on the right hand-side refers to the integrated out momenta and 
the second to the total excess potential energy $\Uexs(\Vhat)$ expressed as a function
of the instantaneous volume as shown in appendix~\ref{app_affineenergy}.

\paragraph*{Non-affine contribution.}
An immediate consequence of the MC-gauge is, of course, that the fluctuations of $\Pidhat$
vanish for the \NVT-ensemble
and that, hence, 
\begin{equation}
\etaFV = \etaFexV \equiv \beta V \left.\la \delta \Pexhat^2 \ra\right|_1
\label{eq_etaPV}
\end{equation}
with the instantaneous excess pressure $\Pexhat$ being computed using Kirkwood's expression, 
eq.~(\ref{eq_PexKirkwood}). According to eq.~(\ref{eq_cmodVP_etaA}) the correlation coefficient 
$\cmodVP$ is thus a function of $\etaFexV/\etaA$ and {\em vice versa}.

\paragraph*{Affine (Born) contribution.}
The task is now to compute the pressure fluctuation in the \NPT-ensemble.
We note first for the \NPT-weight of a configuration at volume $\Vhat$ that 
\begin{eqnarray}
\left[e^{-\beta (\Hams(\Vhat) + P \Vhat)}\right]^{\prime}  & = &
\beta (\Phat - P) e^{-\beta (\Hams(\Vhat) + P \Vhat)} 
\label{eq_partint} \\
\mbox{ where} \  \Phat & \equiv & -\Hams^{\prime}(\Vhat) \ \ \ \mbox{($\Phat$-gauge)}  \label{eq_Phatdef}
\end{eqnarray}
{\em defines} the instantaneous total pressure $\Phat = \Pidhat + \Pexhat$.
Note that this definition is consistent with the MC-gauge, eq.~(\ref{eq_MCgauge}), for $\Pidhat$ and eq.~(\ref{eq_Hams}).
As shown in appendix~\ref{app_volumerescal}, $\Pexhat \equiv - \Uexs^{\prime}(\Vhat)$ is also consistent with the 
Kirkwood excess pressure, eq.~(\ref{eq_PexKirkwood}), for pair potentials.
Using eq.~(\ref{eq_partint}) the second moment $\langle (\Phat - P)^2 \rangle$ can be readily obtained by 
integration by parts. This yields
\begin{equation}
\etaA \equiv \etaFP = 
-\left.V \la \Phat^{\prime}(\Vhat)\ra\right|_0 =
\left.V \la \Hams^{\prime\prime}(\Vhat)\ra\right|_0 
\label{eq_etaA}
\end{equation}
where {\em apriori} the average is understood to be taken over all states $s$ of the system
and all volumes $\Vhat$ at imposed $P$. 
(The boundary terms for the integration by parts over $\Vhat$ can be neglected for sufficiently
large systems since the \NPT-weight for $\Vhat$ gets strongly peaked around $V$.)
It is of importance that the 
{\em fluctuation} $\etaFP$ has thus been reduced to a {\em simple average}. 
This allows its computation more conveniently by \NVT-ensemble simulations. 
Using eq.~(\ref{eq_Hams}) it follows further that $\etaA = \etaAid + \etaAex$ with
\begin{eqnarray}
\etaAid & \equiv & V \left.\la - \kBT N \log(\Vhat)^{\prime\prime} \ra\right|_0 
\approx \Pid \label{eq_etaAid} \\
\etaAex & \equiv & \left.V \la \Uexs^{\prime\prime}(\Vhat)\ra\right|_0  \label{eq_etaAex}
\end{eqnarray}
where we have used for the ideal contribution that for sufficiently large systems 
$V \langle 1/\Vhat^2 \rangle \approx \langle 1/\Vhat \rangle \approx 1/V$.
As seen from the affine excess energy discussed in appendix~\ref{app_affineenergy},
it follows for pair potential interactions that
\begin{equation}
\etaAex = \etaB + \Pex,
\label{eq_etaAexpair}
\end{equation}
i.e. both coefficients $\etaAex$ and $\etaB$ are equivalent. We stress again that $\Pid$, $\Pex$, $\etaB$ and $\etaAex$ are 
simple averages and can thus be evaluated readily in both ensembles using eq.~(\ref{eq_Vhatn}).
Substituting eqs.~(\ref{eq_etaPV},\ref{eq_etaAex},\ref{eq_etaAexpair}) into the transform
eq.~(\ref{eq_Pfluctutrans}), this finally confirms eq.~(\ref{eq_Rowl}).

\subsection{Correlations at constant $P$}
\label{theo_NPT}
We focus now on the fluctuations of the pressure contributions in the \NPT-ensemble.
According to eq.~(\ref{eq_etaAexpair}) we have $P + \etaB = \etaA \equiv \etaFP$.
If the Rowlinson functional $\Krowl$ is measured at imposed $P$ this implies
\begin{eqnarray}
\KrowlP & = & \etaFP - \etaFexP \nonumber \\
& = & \beta V \left.\la \delta \Pidhat^2 \ra\right|_0 + 2 \beta V\left.\la \delta \Pidhat \delta \Pexhat \ra\right|_0.
\label{eq_KrowlPcorrfunc}
\end{eqnarray}
%
As a next step we demonstrate the relations 
\begin{eqnarray}
\etaFidP 
\equiv \beta V \left.\la \delta \Pidhat^2 \ra\right|_0  
& = & K \ x^2 \label{eq_etaFid} \\
\etaFmixP 
\equiv \beta V \left.\la \delta \Pidhat \delta \Pexhat \ra\right|_0 
& = &  K \ x(1 - x) \label{eq_etaFmix} 
\end{eqnarray}
(with $x = \Pid/K$ being again the scaling variable)
from which eq.~(\ref{eq_key}) is then obtained by substitution into 
eq.~(\ref{eq_KrowlPcorrfunc}).
Remembering eq.~(\ref{eq_MCgauge}), eq.~(\ref{eq_etaFid}) is obtained from the fluctuations
of the inverse volume, eq.~(\ref{eq_VinvfluctuA}).
The relation eq.~(\ref{eq_etaFmix}) describing the coupling of ideal and excess pressure is
implied by eq.~(\ref{eq_VinvfluctuB}) for $n=-1$ and using $\Phat = \Pidhat + \Pexhat$ and eq.~(\ref{eq_etaFid}).
We emphasize that eq.~(\ref{eq_key}) or eq.~(\ref{eq_KrowlPcorrfunc}) 
do not completely vanish for finite $T$, i.e. finite $x$, as does the corresponding
stress fluctuation expression for the shear modulus $G$ at imposed shear stress $\tau$ \cite{WXP13}.
Having thus demonstrated eq.~(\ref{eq_etaFid}) and eq.~(\ref{eq_etaFmix}) 
and using the already stated relation for the total pressure, eq.~(\ref{eq_etaA}),
one confirms finally for the fluctuations of the excess pressure $\Pexhat$ that
\begin{equation}
\etaFexP \equiv \beta V \left.\la \delta \Pexhat^2 \ra\right|_0 = \etaAex - K \ x(1 - x) \label{eq_etaFex} 
\end{equation}
with $x= \Pid/K$ being the reduced ideal pressure.
\begin{figure}[t]
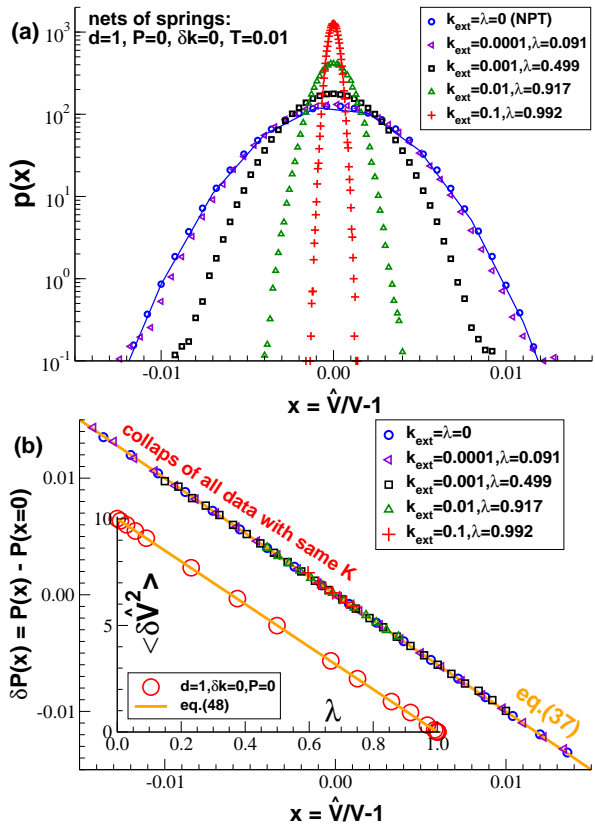

\centerline{\resizebox{0.9\columnwidth}{!}{\includegraphics*{fig1a}}}
\centerline{\resizebox{0.9\columnwidth}{!}{\includegraphics*{fig1b}}}
\caption{Simple 1D nets without noise ($\dk=0$) at temperature $T=0.01$ 
and pressure $P=0$ (mean volume $V=10^3$, number density $\rho=1$)
for different values $\lambda$ of the external spring as indicated:
{\bf (a)} Normalized histograms $p(x)$ of the rescaled instantaneous volume $x = \Vhat/V-1$.
The distributions are Gaussian as shown by the line indicated for
the standard \NPT-ensemble ($\lambda=0$). 
{\bf (b)} Pre-averaged pressure $\delta P(\Vhat)$ traced as a function of the rescaled instantaneous volume $x$. 
The bold line shows eq.~(\ref{eq_PVhat}) for the compression modulus $K=1$ of the system. 
Inset: The mean-squared volume fluctuations $\dVtyplam$ decrease linearly
with $\lambda$ according to eq.~(\ref{eq_dVtypdiff}) or eq.~(\ref{eq_dV_lamB})
as shown by the solid line.
\label{fig_plam}
}
\end{figure}

\subsection{Fluctuations in different $\lambda$-ensembles}
\label{theo_dAdBL}

\paragraph*{Introduction.}
The transformation relations for simple averages and fluctuations between the {\em standard} 
conjugated ensembles have been given in the 1960s by Lebowitz, Percus and Verlet \cite{Lebowitz67}. 
Focusing on the volume $V$ as the only relevant extensive variable and the conjugated (reduced)
pressure $\beta P$ of the system as the only intensive variable \cite{foot_intensive,foot_classical}
we rederive now their results for {\em generalized} $\lambda$-ensembles. 
We illustrate several points made by the histograms presented in fig.~\ref{fig_plam}
for simple 1D nets of harmonic springs as described in sect.~\ref{sec_algo} below.

\paragraph*{Histogram.}
%
Let us assume a normalized distribution $\plam(\Vhat)$ 
with a sharp and {\em symmetric} maximum at the mean volume $V$. 
Examples for such histograms are given in panel (a) of fig.~\ref{fig_plam} for several values 
$\lambda = \Kext/(\Kext + K)$. 
The standard \NPT-ensemble corresponds to the value $\lambda=0$. Note that the volume $V$ of the \NPT-ensemble
at $P=0$ is taken as the reference volume $\Vext$ of the external spring potential. Due to this choice 
neither the mean volume $V$ nor the pressure $P$ do change with increasing $\lambda$,
i.e. all ensembles correspond to the {\em same} thermodynamic state.
All distributions are Gaussian, becoming sharper with increasing $\lambda$. 
(The limit $\lambda \to 1$ corresponds to a Dirac $\delta$-function.)
The width of the distribution around its maximum being characterized by the parameter $\lambda$,
we introduce the notation $\dVtyplam \equiv \langle \delta \Vhat^2 \rangle$ for the mean-squared width.
The width $\dVtyplam$ of simple 1D nets is presented in the inset of panel (b) of fig.~\ref{fig_plam}.

\paragraph*{Observables.}
We write $\Ahat(\Vhat)$ for the expectation value of an observable ${\cal A}$
at a given state $s$ of the (full) phase space of a system at volume $\Vhat$.
Let us first focus on the average $A(\Vhat)$ of $\Ahat(\Vhat)$ taken over all states $s$.
As an example we show in panel (b) of fig.~\ref{fig_plam} the mean pressure $P(\Vhat)$ averaged over
all systems $s$ of a $\lambda$-ensemble found at a given instantaneous volume $\Vhat$. 
As one expects, $P(\Vhat)$ is seen to decrease linearly to leading order,
\begin{equation}
\delta P(\Vhat) \equiv P(\Vhat) - P(V) \approx - \frac{K}{V} \delta \Vhat, \ \delta \Vhat \equiv \Vhat -V,
\label{eq_PVhat}
\end{equation}
as stressed by the bold line. 
Importantly, $P(\Vhat)$ does {\em not} depend on $\lambda$.
(As one expects, the statistics deteriorates for $(\Vhat-V)^2/\dVtyplam \gg 1$.)
More generally, a Taylor expansion of $A(\Vhat)$ around $V$ yields 
\begin{equation}
A(\Vhat) = a_0 + a_1 \delta \Vhat + \frac{1}{2} a_2 \delta \Vhat^2 + \ldots
\label{eq_AVhatexpand}
\end{equation} 
where a coefficient $a_n$ denotes the $n$th derivative of $A(\Vhat)$ with respect to $\Vhat$ taken at $\Vhat=V$.

\paragraph*{Simple averages.}
The average taken over all properly weighted volumes of the investigated $\lambda$-ensemble 
thus becomes to leading order
\begin{equation}
\la A(\Vhat) \ra = \int \ddiff \Vhat \plam(\Vhat) A(\Vhat) \approx 
a_0 + \frac{a_2}{2} \dVtyplam 
\label{eq_avAlam}
\end{equation}
with $a_0 = A(V)$. 
We have used here that $\plam(\Vhat)$ is symmetric around $V$ which implies that the linear term in
eq.~(\ref{eq_AVhatexpand}) must drop out. The difference 
of a simple average 
taken at $\lambda < 1$ and taken in the \NVT-ensemble ($\lambda=1$) is then given by the second derivative $a_2$ times the 
mean-squared width $\dVtyplam$ which increases at most linearly with $V$ for $\lambda=0$. 
Since $a_2$ does not depend on the ensemble, the correction must decay {\em at least as fast} as for the \NPT-ensemble
discussed in the literature \cite{Lebowitz67}.
Hence, simple averages computed for any ensemble with $0 \le \lambda \le 1$ become rapidly indistinguishable.

\paragraph*{Averaged fluctuations.}
As a next step let us consider the correlation function
$\langle \delta A \delta B \rangle = \langle A B \rangle - \langle A \rangle \langle B\rangle$
of the observables $A(\Vhat)$ and $B(\Vhat)$. Using again the symmetry of the distribution 
$\plam(\Vhat)$ around $V$ it is seen that
\begin{equation}
\langle \delta A \delta B \rangle 
= \underline{\langle a_0 b_0 \rangle - \langle a_0 \rangle \langle b_0 \rangle} + a_1 b_1 \dVtyplam
\label{eq_dAdB_A}
\end{equation}
where the coefficients $a_n$ and $b_n$ stand for the respective derivatives of $A(\Vhat)$ and $B(\Vhat)$ taken at $\Vhat=V$
used for the Taylor expansion around $V$. The underlined term vanishes since $a_0 = A(V)$ and $b_0 = B(V)$ are constants.
Note also that $\langle \delta A \delta B \rangle \to 0$ for $\lambda \to 1$.
This is due to the fact that we have correlated here the {\em pre-averaged} observables $A(\Vhat)$ and $B(\Vhat)$
instead of the expectation values $\Ahat$ and $\Bhat$ which depend not only on the volume $\Vhat$ but also
on the state $s$ of the system. Replacing thus $A \to \Ahat$, $B \to \Bhat$, $a_n \to \hat{a}_n$ and $b_n \to \hat{b}_n$
we have thus in addition to average over all possible states and the underlined term in eq.~(\ref{eq_dAdB_A})
remains thus finite
\begin{equation}
\langle \dAhat \dBhat \rangle 
= \underline{\langle \delta \hat{a}_0 \delta \hat{b}_0 \rangle} + \langle \hat{a}_1 \hat{b}_1 \rangle \dVtyplam.
\label{eq_dAdB_AA}
\end{equation}
Since $\langle \hat{a}_1 \hat{b}_1 \rangle$ is a simple average, this result
can be reformulated using a more natural notation as 
\begin{equation}
\left.\la \dAhat \dBhat \ra\right|_{\lambda}  = \left.\la \dAhat \dBhat \ra\right|_{1} 
 + \frac{\partial A}{\partial V} \frac{\partial B}{\partial V} \ \dVtyplam
\label{eq_dAdB_B}
\end{equation}
where $|_{\lambda}$ indicates that the average is taken over properly weighted volumes in a general $\lambda$-ensemble
and $|_1$ indicates the \NVT-average ($\lambda=1$) taken at the maximum of the distribution $\plam(\Vhat)$.
Using eq.~(\ref{eq_dAdB_B}) for the \NPT-limit ($\lambda=0$) this is seen to be identical to eq.~(2.11) 
given in ref.~\cite{Lebowitz67}. Substracting this reference from the general $\lambda$ case yields 
\begin{equation}
\left.\la \dAhat \dBhat \ra\right|_{\lambda} = \left.\la \dAhat \dBhat \ra\right|_{0} 
 -  \left(\dVtypsq - \dVtyplam \right) \frac{\partial A}{\partial V} \frac{\partial B}{\partial V}.
\label{eq_dAdB_C}
\end{equation}
Equation~(\ref{eq_dAdB_C}) can be further simplified using
\begin{equation}
\dVtypsq - \dVtyplam = \frac{V}{\beta K} - \frac{V}{\beta (K+ \Kext)} = \frac{V \lambda}{\beta K}
\label{eq_dVtypdiff}
\end{equation}
for the difference of the volume fluctuations in both ensembles. We have used here eq.~(\ref{eq_K_NPT})
for the \NPT-ensemble and the fact that for general $\lambda$ the external spring is {\em parallel} to the system, 
i.e. the effective modulus must be the sum of the system modulus $K$ and spring modulus $\Kext$.
That eq.~(\ref{eq_dVtypdiff}) holds is confirmed by the data presented in the inset of panel (b) of fig.~\ref{fig_plam}.
Following ref.~\cite{Lebowitz67} we also rewrite the derivatives with respect
to the volume $V$ as derivatives with respect to the pressure $P$ of the system
\begin{equation}
\frac{\partial A}{\partial V} \frac{\partial B}{\partial V} = \left(\frac{\partial P}{\partial V} \right)^2 \frac{\partial A}{\partial P} \frac{\partial B}{\partial P}
= \left( \frac{K}{V} \right)^2 \frac{\partial A}{\partial P} \frac{\partial B}{\partial P}
\label{eq_AB_VtoP}
\end{equation}
where we have used $\partial P / \partial V = - K/V$ in the last step.
Using the above three equations this yields finally 
\begin{equation}
\left.\langle \dAhat \dBhat \rangle\right|_{\lambda} =
\left.\langle \dAhat \dBhat \rangle\right|_{0} -
\frac{K \lambda}{\beta V} \ \frac{\partial A}{\partial P} \frac{\partial B}{\partial P},
\label{eq_dAdB_lam}
\end{equation}
which compares the correlations in a general $\lambda$-ensemble ($0 \le \lambda \le 1$) with 
the correlations in an \NPT-ensemble ($\lambda=0$). 
Note that for $\lambda=1$ this is consistent with the original transformation, eq.~(\ref{eq_dAdB}), 
derived in ref.~\cite{Lebowitz67}.

\subsection{Correlations for generalized $\lambda$-ensembles}
\label{theo_corrL}
%
%
Using eq.~(\ref{eq_dAdB_lam}) we restate first several correlations given above where at least one of the observables 
$\Ahat$ and $\Bhat$ is a function of the instantaneous volume $\Vhat$.
Since for $\lambda < 1$ volume fluctuations are not (completely) suppressed, eq.~(\ref{eq_dAdBnoV}) 
cannot be generalized. Instead the already demonstrated results for $\lambda=0$ or $\lambda=1$ are used.
%
For $\Ahat = \Bhat = \Vhat$ this yields, e.g.,
\begin{eqnarray}
\left.\langle \delta \Vhat^2 \rangle\right|_{\lambda} & = & 
\left.\langle \delta \Vhat^2 \rangle\right|_0 - \frac{K\lambda}{\beta V} \left(\frac{\partial V}{\partial P} \right)^2 
\label{eq_dV_lamA} \\
& = & \dVtyp^2 \ (1 - \lambda) \label{eq_dV_lamB} \\
& = & \frac{\kBT V}{K + \Kext} \label{eq_dV_lamC}
\label{eq_dV_lam}
\end{eqnarray}
restating thus eq.~(\ref{eq_dVtypdiff}). 
%
More generally, one confirms for $\Ahat = \Bhat = \Vhat^n$ that
\begin{equation}
\left.\langle \delta (\Vhat^n)^2 \rangle\right|_{\lambda}  = 
n^2 V^{2n-2} \ \dVtyp^2 \  (1 - \lambda). 
\label{eq_dVn_lam}
\end{equation}
The latter result is consistent with eq.~(\ref{eq_VinvfluctuAA}) which is thus shown to
hold for all $\lambda < 1$.
%
%
For $\Ahat = \Vhat$ and $\Bhat = \Phat$ it is seen that
\begin{equation}
-\beta \left.\la \delta \Vhat \delta \Phat \ra\right|_{\lambda} =
-\beta \left.\la \delta \Vhat \delta \Phat \ra\right|_0 + \frac{K \lambda}{V} \frac{\partial V}{\partial P}
= 1-\lambda
\label{eq_VPcorr_lam}
\end{equation}
generalizing thus eq.~(\ref{eq_VPcorr}). More generally, one sees for $\Ahat =  \Vhat^n$ and $\Bhat = \Phat$ that
\begin{equation}
-\beta \left.\la \delta (\Vhat^n) \delta \Phat \ra\right|_{\lambda}
 =  n V^{n-1}  \  (1-\lambda) \label{eq_VinvfluctuB_lam} 
\end{equation}
as expected from eq.~(\ref{eq_VinvfluctuB}). One confirms using eq.~(\ref{eq_VinvfluctuB_lam})
that eq.~(\ref{eq_VinvfluctuBB}) must hold for all $\lambda < 1$.
%
%
Interestingly, the already mentioned linear regression formula $\KmodVP$ 
is seen using eq.~(\ref{eq_dV_lamB}) and eq.~(\ref{eq_VPcorr_lam}) to become 
\begin{equation}
\left.\KmodVP\right|_{\lambda}  \equiv 
\left.-V \la \dVhat \dPhat \ra\right|_{\lambda} / \left.\la \dVhat^2 \ra\right|_{\lambda} = \frac{\kBT V}{\dVtyp^2} = K
\label{eq_KmodVP_lam}
\end{equation}
independent of the ensemble used. 
As an alternative to the strain fluctuation relation eq.~(\ref{eq_dV_lamC}), 
this allows thus the determination of $K$ for all $\lambda < 1$.
The dimensionless correlation coefficient $\cmodVP$ associated to $\KmodVP$ depends however on $\lambda$
\begin{equation}
\cmodVPL  
= \sqrt{\frac{K}{\etaA}} \ \sqrt{\frac{1-\lambda}{1-\lambda K/\etaA}}
\label{eq_cmodVP_lam}
\end{equation}
where we have used eq.~(\ref{eq_Pfluctutrans_lam}) demonstrated below.
Note that $K/\etaA=1$ implies (as before) $\cmodVPL=1$ for all $\lambda < 1$.
For $K/\etaA<1$ the correlation coefficient decreases continuously from its maximum $\sqrt{K/\etaA}$ 
at $\lambda=0$ to zero for $\lambda=1$. As one would expect, this shows that the more the volume fluctuations 
are suppressed by the external constraint, the more $\Vhat$ and $\Phat$ must decorrelate.
%
%
For the transformation of the total pressure auto-correlation $\etaF \equiv \beta V \langle \delta P^2 \rangle$, 
eq.~(\ref{eq_dAdB_lam}) simply implies that
\begin{equation}
\etaFL = \etaFP - K \lambda = \etaA - \frac{K \Kext}{K + \Kext}.
\label{eq_Pfluctutrans_lam}
\end{equation}
Since $\etaA = \Pid + \etaAex$ is a simple average computable in any ensemble,
eq.~(\ref{eq_Pfluctutrans_lam}) may be also used for the determination of $K$.
Using again the MC-gauge and that $\Phat = \Pidhat + \Pexhat$, it is seen from eq.~(\ref{eq_Pfluctutrans_lam})
that the Rowlinson functional $\KrowlL \equiv P + \etaB - \etaFexL$ should become
\begin{equation}
\KrowlL = K \lambda + \etaFidL + 2 \etaFmixL
\label{eq_KrowlPcorrfunc_lam}
\end{equation}
generalizing thus the Rowlinson formula, eq.~(\ref{eq_Rowl}), for $\lambda=1$ 
(remembering that $\etaFid$ and $\etaFmix$ must vanish in the \NVT-ensemble) and 
eq.~(\ref{eq_KrowlPcorrfunc}) for $\lambda=0$.
Considering as a next step the fluctuations of the ideal pressure $\etaFid$ in the MC-gauge,
eq.~(\ref{eq_dVn_lam}) implies that
\begin{equation}
\etaFidL 
\equiv \beta V \left.\la \delta \Pidhat^2 \ra\right|_{\lambda}
= K \ x^2 \ (1-\lambda). \label{eq_etaFid_lam}
\end{equation}
Since $\Phat = \Pid + \Pex$ the latter result together with eq.~(\ref{eq_VinvfluctuB_lam}) allows to
generalize eq.~(\ref{eq_etaFmix_lam}) for the correlations between ideal and excess pressure
contributions. This shows that
\begin{equation}
\etaFmixL \equiv \beta V \left.\la \delta \Pidhat \Pexhat \ra\right|_{\lambda}
=  K \ x(1 - x) \ (1-\lambda).
\label{eq_etaFmix_lam}
\end{equation}
Our central result eq.~(\ref{eq_KrowlL}) announced in the Introduction is thus simply obtained
by substituting  eq.~(\ref{eq_etaFid_lam}) and eq.~(\ref{eq_etaFmix_lam}) into eq.~(\ref{eq_KrowlPcorrfunc_lam}).
Finally, using eq.~(\ref{eq_Pfluctutrans_lam}) together with eq.~(\ref{eq_etaFid_lam}) and 
eq.~(\ref{eq_etaFmix_lam}) one verifies for the fluctuations of the excess pressure fluctuations that
\begin{equation}
\etaFexL = \etaA - \KrowlL 
\label{eq_etaFex_lam}
\end{equation}
as obvious from eq.~(\ref{eq_KrowlL}). The latter result reduces to the Rowlinson expression eq.~(\ref{eq_Rowl}) 
for $\lambda=1$ and using $\etaA=\etaAex+Kx$ to eq.~(\ref{eq_etaFex}) for $\lambda=0$.
\section{Some algorithmic details}
\label{sec_algo}

\paragraph*{Introduction.}
In order to check our predictions we sampled by MC and MD simulation \cite{AllenTildesleyBook,FrenkelSmitBook} 
various model systems for solids and glass-forming liquids in $d=1$, $2$ and $3$ dimensions.
Periodic boundary conditions are used and all systems are first kept at constant pressure $P$ 
using standard barostats \cite{AllenTildesleyBook} as specified below. After equilibrating
and sampling in the \NPT-ensemble, the volume fluctuations are suppressed either by imposing
$V=\Vhat$ or by means of a {\em finite} spring potential. 
Various simple averages, such as the Born-Lam\'e coefficient $\etaB$, and fluctuations, such as the excess 
pressure fluctuation $\etaFex$ or the Rowlinson functional $\Krowl$, are compared  
for all available $\lambda$.  As expected, all simple averages are found within numerical 
accuracy to be identical. 
As discussed in sect.~\ref{sec_simu}, fluctuations are found to transform following the predictions 
based on eq.~(\ref{eq_dAdB}) and eq.~(\ref{eq_dAdB_lam}).
Specifically, we have verified that eqs.~(\ref{eq_VinvfluctuAA},\ref{eq_VinvfluctuBB},\ref{eq_VPcorr_lam})
hold to high precision for all $\lambda$.
About $N=10^4$ particles are typically used.
%
Lennard-Jones (LJ) units are used throughout this work and $\kB$ is set to unity.

\paragraph*{One-dimensional spring model.}
As already seen in fig.~\ref{fig_plam}, the bulk of the presented numerical results has been obtained 
by MC simulation of permanent nets of ideal harmonic springs in strictly $d=1$ dimension \cite{foot_ergodicity}. 
We use a potential energy 
\begin{equation}
U = \frac{1}{2} \sum_l \kl (\xl - \Rl)^2
\label{eq_netspring}
\end{equation}
with $\xl$ being the distance between the connected particles.
The reference length $\Rl$ of the springs is assumed to be constant, $\Rl = R = 1$,
and the spring constants $\kl$ are taken randomly from a uniform distribution of half-width $\dk$ 
centered around a mean value also set to unity. 
We note for later reference that this implies
\begin{equation}
\frac{1}{\la 1/\kl \ra} = \frac{2\dk}{\log((1+\dk)/(1-\dk))}
\label{eq_Kdk}
\end{equation}
as indicated by the bold line in fig.~\ref{fig_K_dk}.
%
Only simple networks are presented here where two particles
$i-1$ and $i$ along the chain are connected by {\em one} spring $l=i$, i.e. all forces $\fl$ 
along the chain are on average identical. 

\paragraph*{Glass-forming liquids.}
Our two-dimensional (2D) systems are polydisperse Lennard-Jones (pLJ) beads as described in ref.~\cite{WXP13}. 
The reported three-dimensional (3D) systems refer to binary Kob-Andersen (KA) mixtures \cite{Kob95} 
sampled by means of MD simulations taking advantage of the LAMMPS implementation \cite{LAMMPS}. 
Starting from the liquid limit well above the glass transition temperature $\Tglass$,
both system classes have been quenched \cite{WXP13}
deep into the glassy state at very low temperatures $T \ll \Tglass$ as may be seen from fig.~\ref{fig_K_T}
\cite{foot_trunc}.

\paragraph*{Barostats.}
The results reported for $d=1$ and $d=2$ have been obtained using local MC moves for the particles and 
global MC moves for the barostat \cite{WXP13}. As described elsewhere \cite{AllenTildesleyBook,WXP13}, 
an attempted volume change $\delta \Vhat = \Vhat_\text{new} - \Vhat_\text{old}$ is accepted if 
$\xi \le \exp(-\beta \delta G)$ with $\xi$ denoting a uniformly distributed random variable 
with $0 \le \xi < 1$ and
\begin{equation}
\delta G = \delta E + \delta \Vhat P - \kBT N \log(\Vhat_\text{new} / \Vhat_\text{old}).
\label{eq_Pimposed}
\end{equation}
The first term $\delta E$ stands for the energy difference associated with the affine displacements
of all particles, the second term $\delta \Vhat P$ imposes the pressure $P$ and
the logarithmic contribution corresponds to the change of the translational entropy,
i.e. the change of the integrated out momentum contribution discussed above, eq.~(\ref{eq_VNfactor}).
While a broad range of pressures has been sampled for the 1D nets, the 2D pLJ beads have been kept at 
only one pressure, $P=2$, for which a glass-transition temperature $\Tglass \approx 0.26$ has been determined \cite{WXP13}.

For more general $\lambda$-ensembles the increment $\delta \Uext(\Vhat)$ of the external spring defined in 
eq.~(\ref{eq_Uext}) is simply added to $\delta G$. 
We assume throughout this work that $\Vext \equiv V$ for the reference volume of the external spring 
with $V$ being the mean volume of an \NPT-ensemble of a given pressure $P$. Due to the symmetry of the 
fluctuations around $V$ for the system and the external spring, this is sufficient to keep this pressure 
constant for all $\lambda$ \cite{foot_generalensemble}. 
%

For our MD simulations of the KA model we have used the Nos\'e-Hoover barostat (``fix npt command")
provided by the LAMMPS code \cite{LAMMPS}.
Following Kob and Andersen \cite{Kob95} a constant pressure $P=1$ has been imposed for all temperatures.
This choice corresponds to glass-transition temperature $\Tglass \approx 0.4$ \cite{Kob95,WXP13}.
%

\section{Computational results}
\label{sec_simu}

As already noted above, various properties have been compared for different boundary conditions 
as characterized by the parameter $\lambda$ while keeping the system at the same state point.
We focus first on the comparison of \NPT- ($\lambda=0$) and \NVT-ensembles ($\lambda=1$)
before we turn in sect.~\ref{simu_lambda} to the more general $\lambda$-ensembles.

\begin{figure}[t]
\centerline{\resizebox{0.9\columnwidth}{!}{\includegraphics*{fig2}}}
\caption{Compression modulus $K$ for 1D nets at temperature $T=0.01$ and pressure $P=0$
computed using the rescaled volume fluctuations $\KdvolP$ (filled spheres),
the Rowlinson stress fluctuation formula $\KrowlV$ (crosses),
the difference between the total pressure fluctuations in both ensembles (squares) and
the fluctuations of the inverse volume $\Pid^2/\etaFidP$ (large spheres).
The compression modulus decreases strongly with $\dk$ 
(solid line). Also indicated are the ``affine" contribution $\etaFP = 1 + 2 \Pid$ 
to the compression modulus, eq.~(\ref{eq_etaAnet}), and the ``non-affine" correction $\etaFV$ 
which is seen to increase with $\dk$ according to eq.~(\ref{eq_etaFexnet}) as indicated by the dash-dotted line.
\label{fig_K_dk}
}
\end{figure}

\begin{figure}[t]
\centerline{\resizebox{0.9\columnwidth}{!}{\includegraphics*{fig3}}}
\caption{Compression modulus $K$ {\em vs.} temperature $T$ comparing again
the values obtained using $\KdvolP$, $\KrowlV$, eq.~(\ref{eq_Pfluctutrans}) 
and eq.~(\ref{eq_etaFid}).
The upper data refer to the results obtained for the KA model in 3D \cite{Kob95,WXP13}
which have been shifted upwards (factor 4) for clarity, 
the data in the middle to systems of glass-forming 2D pLJ beads at $P=2$ \cite{WXP13} and
the lower data to simple 1D nets of harmonic springs at $P=0$ with identical spring constants ($\dk=0$).
\label{fig_K_T}
}
\end{figure}

\begin{figure}[t]
\centerline{\resizebox{0.9\columnwidth}{!}{\includegraphics*{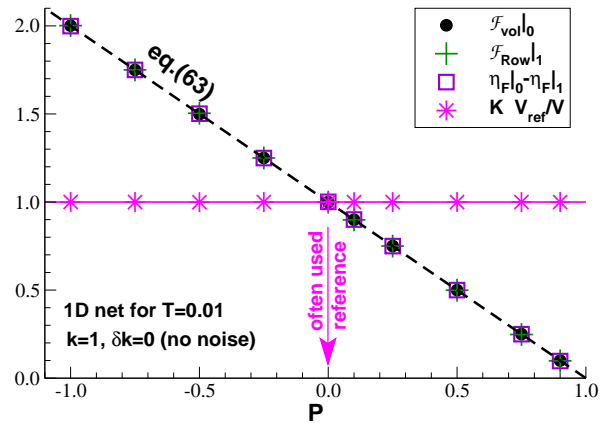}}}
\caption{Compression modulus $K$ {\em vs.} pressure $P$ for 1D nets
without noise ($\dk=0$) at $T=0.01$.
The compression modulus 
decreases with $P$ simply since the volume $V$ of the system linearly decreases
as $V= \Vref (1 - P/\Kref)$ with $\Kref$ being the compression modulus
at $P=0$ where $V=\Vref$. If $K$ is rescaled with $\Vref/V$ (spheres) 
this yields a pressure independent material constant $\Kref$.
\label{fig_K_P}
}
\end{figure}

\subsection{Compression modulus $K$}
\label{simu_K}

\paragraph*{Comparing different methods.}
As shown in figs.~\ref{fig_K_dk}-\ref{fig_K_P}, the compression modulus $K$ may be determined either using 
the volume fluctuations in the \NPT-ensemble, eq.~(\ref{eq_K_NPT}), or using Rowlinson's stress fluctuation formula, 
eq.~(\ref{eq_Rowl}), for the \NVT-ensemble (crosses). 
The same values of $K$ are obtained from the transform of the pressure fluctuations, 
eq.~(\ref{eq_Pfluctutrans}), and from the ideal pressure fluctuations $\etaFidP$, eq.~(\ref{eq_etaFid}), 
as indicated by the large spheres which thus confirms both relations.

\paragraph*{Harmonic spring networks.}
As seen in fig.~\ref{fig_K_dk}, the compression modulus of the 1D nets decreases strongly with $\dk$.
To understand the scaling indicated by the bold line let us consider a chain of harmonic strings.
Since the average force acting on each spring
is given by the imposed pressure $P = - \fext/A$, this implies an average length 
$\langle \xl \rangle = R  - P A/\kl$ for a spring constant $\kl$
and an average volume $V = A \sum_l \langle \xl \rangle$. A pressure increment $\delta P$
thus leads to a volume change $\delta V = - A \delta P \sum_l 1/\kl$. For the compression 
modulus $K = - V \delta P/\delta V$ at pressure $P$ this yields 
\begin{equation}
K = \Kref - P \mbox{ with } \Kref = \rhoref / \langle 1/\kl \rangle
\label{eq_Knet}
\end{equation}
being the compression modulus of the unstressed reference system at $P=0$
and $\rhoref = 1/AR$ the corresponding density.
The compression modulus $K=\Kref$ at zero pressure is thus inversely proportional to
the average inverse spring constant. For our uniformally distributed spring constants
eq.~(\ref{eq_Kdk}) thus implies that $K$ must vanish as a cusp-singularity for 
$\dk \to 1$ \cite{foot_cusp}.
Also indicated in fig.~\ref{fig_K_dk} are the ``affine" contribution $\etaFP$ to $K$
(further discussed in sect.~\ref{simu_affine}) 
and the ``non-affine" contribution $\etaFV$ which is seen to increase with $\dk$.
The decrease of $K$ is thus due to the increase of the non-affine contribution.
%

\paragraph*{Temperature dependence.}
Figure~\ref{fig_K_T} shows the temperature dependence of the compression modulus for our three model systems.
As one would expect, $K$ decreases with $T$ for the bead systems kept at same pressure reaching 
the ideal gas compressibility $K = \Kid = \Pid = P$ for large temperatures.
As predicted by eq.~(\ref{eq_Knet}) for strictly harmonic spring chains, the compression modulus $K$ is found 
$T$-independent for all 1D networks and this irrespective on the values $\dk$ and $P$ sampled (not shown).

\paragraph*{Pressure dependence.}
The pressure dependence of the compression modulus for 1D nets is investigated in fig.~\ref{fig_K_P}.
The dashed line indicating the linear behavior predicted by eq.~(\ref{eq_Knet}) for chains of harmonic springs
perfectly fits the measured data points.
A comment is in order here:
Since the springs are permanently fixed and ideal, the {\em intrinsic} mechanical properties of the systems do not change 
with the external load. In fact, the second derivative of the free energy $F(T,V)$ with respect to the volume $V$
becomes constant for these highly idealized systems
\begin{equation}
\frac{K}{V} \equiv \frac{\partial^2 F(T,V)}{\partial V^2} 
= \text{const} = \frac{\Kref}{\Vref}
\label{eq_Kref}
\end{equation}
with $\Vref$ being the reference volume of the system at zero pressure. The difference between the thermodynamic compression
modulus $K$ and the constant $\Kref$ simply arises since in all thermodynamic relations, such as eq.~(\ref{eq_Kdef}),
the volume $V$ of the current state is taken to make the modulus system-size independent (intensive) and {\em not} a
reference volume at a certain pressure $P$. Since for an idealized elastic body as our 1D nets 
$V = \Vref ( 1 - P/\Kref)$, this implies the linear relation $K = \Kref -P$ (dashed line). 
Note that for solids the applied pressures are for once small compared to the moduli and 
the systems can be taken linearly and without (or at least with negligible) plastic rearrangement 
from a zero-pressure reference to a finite $P$.
It is thus possible to define elastic material properties such as $\Kref$ which are {\em independent} 
of the applied external stresses. If possible, this is a conceptionally and practically very 
useful separation of conditions and effects. However, it is obviously pointless to take the zero-pressure 
volume (or the volume of any other pressure) as a reference for a colloidal or polymeric system from 
which other pressure states are described by a linear extrapolation. 
Please note that being in principle not very deep (being a matter of definitions), 
this issue creates significant confusion between people from different communities.
See appendix~\ref{app_generalmodul} for addition comments on this issue.

\begin{figure}[t]
\centerline{\resizebox{0.9\columnwidth}{!}{\includegraphics*{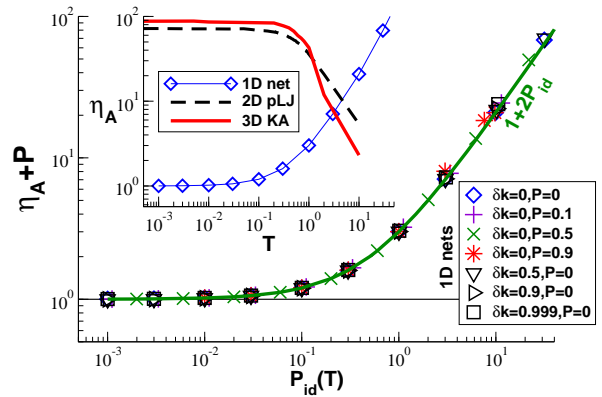}}}
\caption{Affine dilatational elasticity $\etaA$ determined from the pressure fluctuations
$\etaFP$ in the \NPT-ensemble.
Inset: $\etaA$ as a function of $T$ for 1D nets ($\dk=0$, $P=0$),
2D pLJ beads ($P=2$) and 3D KA binary mixtures ($P=1$).
Main panel: Data collapse for $\etaA + P$ as a function of $\Pid(T)$ for 1D nets
for a broad range of parameters as indicated. The bold line corresponds to the expected
behavior for harmonic springs.
\label{fig_affine}
}
\end{figure}

\subsection{Affine compressibility contribution}
\label{simu_affine}

%
The inset of fig.~\ref{fig_affine} presents for our three model systems the correlation function 
$\etaF \equiv \beta V \langle \delta \Phat^2 \rangle$ of the total pressure in the \NPT-ensemble 
measured as a function of temperature while keeping the mean pressure $P$ constant.
As shown in sect.~\ref{sec_theo}, the ``affine dilatational elasticity" $\etaA \equiv \etaFP$ 
yields the leading contribution to the compression modulus $K$.
We note first that $\etaA$ increases with decreasing temperature for our two glass-forming liquids
in $d=2$ and $d=3$ dimensions (due to the increasing repulsion of the LJ beads) levelling off 
below the glass-transition temperature $\Tglass$ of each model. At contrast, $\etaA$ is seen to increase 
monontonously with temperature. This qualitatively different behavior needs to be explained.
We remind first that $\etaA$ can be expressed by the simple average,
eq.~(\ref{eq_etaA}), which can be evaluated in any ensemble. For the pair potentials
we focus on here this yields $\etaA = \Pex + \etaB = \Pid + \etaAex$ with $\etaB$ 
being the Born-Lam\'e coefficient, eq.~(\ref{eq_etaB}),
as we have explicitly checked for all models \cite{foot_trunc}.
%
%
%
Using eq.~(\ref{eq_etaAexdimone}) one sees that for harmonic springs in $d=1$ dimensions 
\begin{equation}
\etaA = \Pid +  \frac{1}{V} \sum_l \kl \la \xl^2 \ra
\label{eq_etaA_A}
\end{equation}
for a given quenched realization of spring constants $\kl$.
Since in the \NPT-ensemble each spring is uncorrelated, we can use
that $\langle \xl^2 \rangle - \langle \xl \rangle^2 =  \kBT/\kl$
for the thermal fluctuation of every spring $l$.
This implies that $\etaA = 2\Pid +  \sum_l \kl \la \xl \ra^2/V$.
Since the average length of a spring at constant pressure $P$ is given by $\langle \xl \rangle = R - P A/\kl$ and 
since the mean volume $V$ is the sum of averaged lengths $\langle \xl \rangle$ times the surface $A$, 
it follows using eq.~(\ref{eq_Knet}) for the compression modulus $K$ that
\begin{equation}
\etaA = 2 \Pid + \Kref - P + \Kref \ \frac{\la \kl \ra \la 1/\kl \ra - 1}{1 - \la 1/\kl \ra P A /R }
\label{eq_etaAnet}
\end{equation}
with $\Kref = \rhoref R^2/\langle 1/\kl \rangle$ and $\rhoref = 1/A R$ characterizing the zero-pressure reference.
For systems of identical springs this reduces to $\etaA = 2 \Pid + \Kref -P$.
With $\Kref=1$ this is shown by the bold line in the main panel of fig.~\ref{fig_affine}. For polydisperse systems at zero pressure 
we have instead $\etaA = 2 \Pid + \rhoref R^2 \langle \kl \rangle$ as already shown by 
the dashed line in fig.~\ref{fig_K_dk}. Within the units used this corresponds numerically also to $\etaA = 1 + 2 \Pid$.
Hence, all 1D nets without noise or without applied pressure should collapse on the same master curve
if $\etaA+P$ is plotted as a function of the ideal pressure $\Pid(T)$. 
This is confirmed by the data presented in the main panel of fig.~\ref{fig_affine}.

\begin{figure}[t]
\centerline{\resizebox{0.9\columnwidth}{!}{\includegraphics*{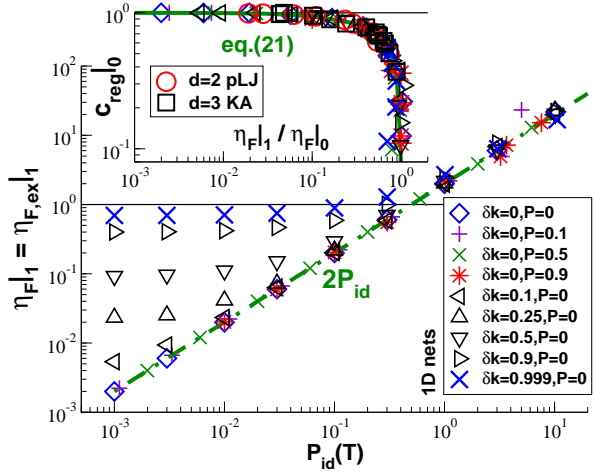}}}
\caption{Characterization of non-affine elastic response.
Main panel: Excess pressure fluctuation $\etaFexV$ as a function of $\Pid(T)$
for 1D nets for different $\dk$ and $P$. 
Systems with no ($\dk=0$) or little polydispersity and all high-temperature systems collapse 
on the dashed line indicating $\etaFexV = 2\Pid$.
Inset: Dimensionless regression coefficient $\cmodVPP$ for all three model systems
as a function of $\etaFV/\etaFP = \etaFexV/\etaA$ confirming eq.~(\ref{eq_cmodVP_etaA}) 
as indicated by the bold line.
\label{fig_nonaffine}
}
\end{figure}

\subsection{Non-affine compressibility contribution}
\label{simu_nonaffine}

The non-affine deviations from the affine Born contribution $\etaA$ to the compression modulus $K$
are given according to the Rowlinson formula, eq.~(\ref{eq_Rowl}), by the fluctuation $\etaFexV$
of the excess pressure in the \NVT-ensemble. Focusing on one low temperature we have already
seen $\etaFexV$ in fig.~\ref{fig_K_dk} as a function of $\dk$.
Plotted as a function of the ideal pressure $\Pid(T)$, the main panel of fig.~\ref{fig_nonaffine} 
presents $\etaFexV$ for various $\dk$ and $P$ as indicated. 
Since $\etaFexV = \etaA - K$, it follows from eq.~(\ref{eq_Knet}) and eq.~(\ref{eq_etaAnet}) characterizing,
respectively, the compression modulus and the affine dilatational elasticity, that for 1D nets
of harmonic springs
\begin{equation}
\etaFexV = 2 \Pid + \frac{1 - 1/ \la 1/\kl \ra}{1 - P \la 1/\kl \ra} 
\label{eq_etaFexnet}
\end{equation}
where we have used that the mean spring constant $\langle \kl \rangle$, 
the reference length $R$ of the springs and the surface $A$ are all arbitrarily set to unity.
Assuming $P=0$, eq.~(\ref{eq_etaFexnet}) is traced in fig.~\ref{fig_K_dk} as a function of $\dk$ (dash-dotted line).
Since the second term vanishes for identical springs, this relation implies $\etaFexV = 2 \Pid$ for all temperatures 
and pressures as shown in the main panel of fig.~\ref{fig_nonaffine} by the dash-dotted
power-law slope. This is confirmed by the data presented for $\dk=0$. 
As one expects, this also gives the high-temperature limit for {\em all} systems. 
Note that $\etaFexV$ only vanishes in the low-$T$ limit of strictly identical springs.
Since the second term in eq.~(\ref{eq_etaFexnet}) is a finite constant for $\dk > 0$,
the corresponding data must thus level off for $T \to 0$. 
Using a simple example we have thus confirmed the more general finding by Lutsko \cite{Lutsko89}
that the non-affine contributions to the elastic moduli need not necessarily vanish in the low-$T$ limit.
Interestingly, due to the $P$-dependent denominator the low-$T$ non-affine contribution
becomes more and more relevant for larger pressures. It becomes thus increasingly
difficult to reach the asymptotic high-$T$ limit (not shown) \cite{foot_etaFexVscaling}.
%
%
As stressed in sect.~\ref{theo_etaFtrans}, the correlation coefficient $\cmodVP$ of the
instantaneous volumes and pressures in the \NPT-ensemble is set by the reduced non-affine 
contribution $\etaFV/\etaFP = \etaFexV/\etaA$.
This relation is verified in the inset of fig.~\ref{fig_nonaffine} presenting for all three studied model system
a perfect collapse of the data on the master curve (bold solid line) predicted by eq.~(\ref{eq_cmodVP_etaA}).
Hence, the coefficient $\cmodVP$ may be seen as a dimensionless and properly normalized ``order parameter"
characterizing the relative importance of the non-affine displacements. 
This also shows that it is equivalent to specify for a system either $(K,\etaA)$, $(\etaA,\etaFexV)$ or $(K,\cmodVP)$.

\begin{figure}[t]
\centerline{\resizebox{0.9\columnwidth}{!}{\includegraphics*{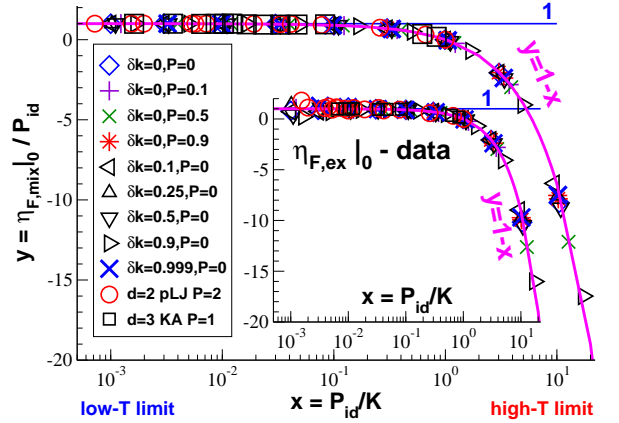}}}
\caption{Stress fluctuations in the \NPT-ensemble.
Large spheres refer to 2D pLJ beads for $P=2$, squares to 3D KA beads for $P=1$,
all other symbols to 1D nets for different $\dk$ and $P$ as indicated.
Main panel:
$y= \etaFmixP/\Pid$ as a function of the reduced ideal pressure $x = \Pid/K$ confirming eq.~(\ref{eq_etaFmix}).
Inset: Similar scaling for the reduced correlation function $y = (\etaAex-\etaFexP)/\Pid$ using the same symbols 
as in the main panel. The bold line represents eq.~(\ref{eq_etaFex}).
\label{fig_etaFmixexP}
}
\end{figure}

\subsection{Correlations in the $\NPT$-ensemble}
\label{simu_NPT}

As shown in fig.~\ref{fig_etaFmixexP}, we have checked the predicted correlations between the
ideal and the excess pressure fluctuations $\etaFmixP$ and the auto-correlations of the
excess pressure $\etaFexP$. To make all investigated models comparable the reduced 
correlation functions $y = \etaFmixP /\Pid$ (main panel) and $y=(\etaAex-\etaFexP)/\Pid$ (inset)
are traced as a function of the reduced ideal pressure $x = \Pid/K$ with $K$ as determined independently above. 
Please note that compared to our theoretical predictions we have scaled the vertical axes of the data 
not with $K$, but using the ideal pressure $\Pid$, i.e. the predictions have been divided by $x$. 
This was done for presentational reasons since thus the low-$T$ asymptote becomes a constant.
A perfect data collapse on the (rescaled) prediction $y = 1-x$ (bold line) is observed for all systems.
As shown in ref.~\cite{errat} a similar data collapse is achieved for the
Rowlinson formula if $y = \KrowlP/\Pid$ is plotted as a function of $x$. 
(Since a related scaling plot is presented in sect.~\ref{simu_lambda},
we do not repeat this figure here.)
Interestingly, the latter scaling does {\em not} depend on the MC-gauge 
which has been used above to simplify the derivation of eq.~(\ref{eq_key}).
Please note that for our glass-forming liquids it is not possible (for the imposed pressures)
to increase $x$ beyond unity ($K \ge \Pid$) and the deviations from the predicted plateau 
$y = \fscalP(x)/x \approx 2$ for $x \ll 1$ are thus necessarily small
and can be easily overlooked \cite{WXP13}.
%

\subsection{Generalized $\lambda$-ensembles}
\label{simu_lambda}
%

%
Having characterized the behavior in the \NPT-ensemble ($\lambda=0$) and the \NVT-ensemble ($\lambda=1$) 
we turn now to the general $\lambda$-ensembles. 
As described at the end of sect.~\ref{sec_algo}, these Gaussian ensembles \cite{Hetherington87,CETT06} 
may be realized by attaching an external harmonic spring, eq.~(\ref{eq_Uext}), parallel to the system.
As we have already seen in fig.~\ref{fig_plam}, this allows to gradually reduce 
the volume fluctuations while keeping constant the average pressure $P$ of the system,
i.e. the thermodynamic (average) state of the system remains unchanged \cite{foot_generalensemble}.
Expressed as a function of $\lambda$,
the mean-squared volume fluctuations decrease linearly 
in agreement with eq.~(\ref{eq_dV_lamB}). Confirming eq.~(\ref{eq_dVn_lam}),
a similar scaling is also found for other moments of the volume (not shown).

\begin{figure}[t]
\centerline{\resizebox{0.9\columnwidth}{!}{\includegraphics*{fig8a}}}
\centerline{\resizebox{0.9\columnwidth}{!}{\includegraphics*{fig8b}}}
\caption{Verification of various predictions for generalized $\lambda$-ensembles
for simple 1D nets:
{\bf (a)} identical springs ($\dk=0$) at temperature $T=0.01$ and pressure $P=0$ 
where $K=1$, $\etaA=1.02$, $\cmodVPP \approx 1$ and $\fscalP=0.0199$ for $x=\Pid/K=0.01$ and
{\bf (b)} polydisperse springs with $\dk=0.9$ at 
$T=1.0$ and $P=0$ where $K=0.6$, $\etaA=3.00$, $\cmodVPP = 0.44$ and $\fscalP=0.53$ for $x=\Pid/K=1.68$.
The lines compare the data with the respective prediction:
the thin solid lines show the compression modulus obtained using the regression relation, eq.~(\ref{eq_KmodVP_lam}),
the dash-dotted lines the transformation relation for the pressure fluctuations $\etaFL$, eq.~(\ref{eq_Pfluctutrans_lam}),
the bold solid lines the key claim, eq.~(\ref{eq_KrowlL}), for the Rowlinson functional $\KrowlL$.
\label{fig_corrL}
}
\end{figure}

%
Focusing on simple 1D nets the verification of several predicted pressure and volume-pressure correlation 
functions is presented in fig.~\ref{fig_corrL}.
The data presented in panel (a) has been obtained for systems with identical spring constants
($\dk=0$) at a low temperature $T=0.01$. Volume and pressure fluctuations are thus highly 
correlated in the \NPT-ensemble, i.e. $\cmodVPP \approx 1$, and the fluctuations
of the excess pressure $\etaFexV$ in the \NVT-ensemble is consequently small. 
The second system presented in panel (b)
corresponds to a polydispersity $\dk=0.9$ and a temperature $T=1$ for which 
$\cmodVPP \approx 0.44$ and $\etaFexV/\etaA \approx 0.8$. 
Note that the vertical axes are made dimensionless by rescaling the data using either the
compression modulus $K$ or the affine dilatational elasticity $\etaA$ obtained for $\lambda=0$.
We stress first that the compression modulus $K$ may be computed irrespective of $\lambda$
by linear regression (spheres) of the measured $(\Vhat,\Phat)$ 
as predicted by eq.~(\ref{eq_KmodVP_lam}). The corresponding dimensionless correlation 
coefficient $\cmodVPL$ indicated by the squares if found to decrease monotonously from its 
maximum at $\lambda=0$ to zero in the \NVT-ensemble. In agreement with eq.~(\ref{eq_cmodVP_lam})
the decay is more sudden for values $K/\etaA$ close to unity, i.e. if the non-affine
contribution is negligible as in panel (a), and becomes more and more gradual for smaller $K/\etaA$
as seen in panel (b).
The linear decay of the total pressure correlation function $\etaFL$ demonstrates that the predicted
transformation relation of the total pressure fluctuations, eq.~(\ref{eq_Pfluctutrans_lam}), indeed 
holds as shown by the dash-dotted lines. Note that in the limit $\lambda \to 1$ we have $\etaFL \to \etaFexV$
due to the MC-gauge used for the instantaneous ideal pressure.
Since for the first system in panel (a) the thermal noise is very week, we have $\etaFL \approx \etaFexL$ 
for all $\lambda$ (not shown). Therefore, $\etaFexL$ is only presented in the second panel.
Our key prediction eq.~(\ref{eq_KrowlL}) for the Rowlinson functional is confirmed by the crosses presenting
$\KrowlL/K$ {\em vs.} $\lambda$. 

\begin{figure}[t]
\centerline{\resizebox{0.9\columnwidth}{!}{\includegraphics*{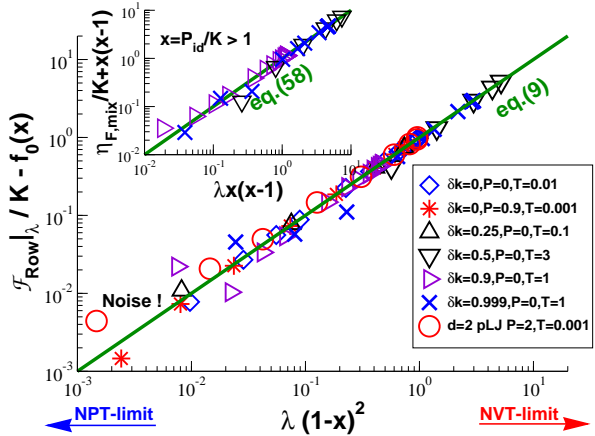}}}
\caption{Scaling with $\lambda$ for various compression moduli $K$ and reduced ideal pressures $x=\Pid/K$.
Main panel: Rescaled Rowlinson functional $\KrowlL$ for different polydispersity $\dk$, 
pressure $P$ and temperature $T$ as indicated. The spheres indicate 2D pLJ beads,
all other symbols data obtained for 1D nets.
Note that the horizontal scaling variable $\lambda (1-\fscalP(x)) = \lambda (1-x)^2 \ge 0$. 
The bold line corresponds to eq.~(\ref{eq_KrowlL}).
Inset: Rescaled data for $\etaFmixL$ as a function of $\lambda x(x-1)$.
The bold bisection-line corresponds to eq.~(\ref{eq_etaFmix_lam}).
Only data for 1D nets with $x>1$ are presented. 
\label{fig_KrowlL}
}
\end{figure}

%
The scaling of $\KrowlL$ is further investigated in the main panel of fig.~\ref{fig_KrowlL} for 
a broad range of 1D nets (with parameters as indicated in the figure) and 2D pLJ beads (spheres) 
for one state point ($T=0.001$, $P=2$). 
According to eq.~(\ref{eq_key},) $\KrowlL/K - \fscalP(x) \to 0$ in the \NPT-limit.
Reshuffling the contributions indicated in eq.~(\ref{eq_key}), the latter limit has been used for the 
vertical axis. Using as scaling variable $\lambda (1 - \fscalP(x)) = \lambda (1-x)^2$ for the
horizontal axis, all data must collapse on the bisection line according to eq.~(\ref{eq_key}).
As may be seen from the double-logarithmic plot, this is indeed the case for a broad range of data sets.
The inset of fig.~\ref{fig_KrowlL} presents a similar scaling plot for the function $\etaFmixL$ characterizing 
the correlations of ideal pressure and excess pressure contributions for 1D nets. The vertical axis corresponds 
again to the prediction of the \NPT-limit, eq.~(\ref{eq_etaFmix}). The horizontal axis $\lambda x (x-1)$ is chosen 
such that according to eq.~(\ref{eq_etaFmix_lam}) all rescaled $\etaFmixL$ have to fall on the bisection line. 
(Since the horizontal axis is logarithmic, only data with $x > 1$ can be represented and, 
hence, only data for 1D nets are given.) 
This is confirmed by the presented data. 

\section{Conclusion}
\label{sec_conc}
%

%
We have revisited in this paper theoretically and numerically various correlations of the normal 
pressure $\Phat$ and its contributions $\Pidhat$ and $\Pexhat$ for isotropic solids and fluids 
using simple coarse-grained models \cite{foot_classical}, e.g., strictly 1D networks of permanently 
fixed springs or the KA model for binary mixtures in three dimensions \cite{Kob95}.
Making more precise several statements made in the appendix of ref.~\cite{WXP13}
and extending the brief communication ref.~\cite{errat}, 
we have compared fluctuations in generalized $\lambda$-ensembles where the volume fluctuations are tuned
by means of an external harmonic spring potential allowing to switch gradually 
between the standard \NPT- ($\lambda=0$) and \NVT-ensembles ($\lambda=1$).
We have stressed that the widely used stress fluctuation formula, eq.~(\ref{eq_Rowl}), 
for the compression modulus $K$ in the \NVT-ensemble may be obtained directly 
\begin{itemize}
\item
without the affine volume rescaling trick for the \NVT-ensemble (reminded in appendix~\ref{app_volumerescal}) 
used first for liquids by Rowlinson \cite{RowlinsonBook} and,
\item
more importantly, without assuming a reference position for the particles 
and a microscopic displacement field which is only possible for solids \cite{Hoover69}
\end{itemize}
using the general thermodynamic transformation rules between conjugated ensembles 
\cite{Lebowitz67} and assuming the systems to be sufficiently large ($V \to \infty$) 
for the given temperature $T$ and compression modulus $K$ \cite{foot_critical}. 
The direct thermodynamic derivation can readily be adapted to the shear modulus $G$ in 
isotropic systems \cite{WXP13} and to the more general elastic moduli characterizing 
anisotropic solids as reminded in appendix~\ref{app_generalmodul} \cite{foot_critical}. 
%
%

The Rowlinson stress fluctuation functional $\Krowl$ \cite{AllenTildesleyBook} 
has been computed deliberately in the unusual \NPT-ensemble ($\lambda=0$) to make manifest 
the general transform eq.~(\ref{eq_Pfluctutrans}) at the heart of the stress fluctuation formalism.
We have demonstrated that $\KrowlP = \Pid (2  - \Pid/K)$, 
i.e. $\KrowlP$ vanishes in the low-temperature limit.
More generally, we have investigated $\KrowlL$ and other correlation functions as a function
of the parameter $\lambda$ characterizing the volume fluctuations. As announced in the Introduction,
eq.~(\ref{eq_KrowlL}), the Rowlinson functional is found to interpolate linearly between the classical 
ensemble limits (fig.~\ref{fig_corrL}). 
Note that the specification of the compression modulus $K$ and the dimensionless
regression coefficient $\cmodVPP$ implies $\etaA$ and $\etaFexV$ (and {\em vice versa}) and 
together with the ideal pressure $\Pid$ (which implies $x=\Pid/K$) this allows the complete 
description of all the discussed correlation functions at different $\lambda$.
%
%
%
Our theoretical and numerically results, especially eq.~(\ref{eq_KrowlL}), may allow to readily calibrate 
(correctness, convergence and precision) the various barostats commonly used \cite{AllenTildesleyBook,LAMMPS}.
%
In the near future we plan 
{\em (i)} 
to also consider systems close to a first-order (e.g., solid to liquid) phase transition 
generalizing the work by Hetherington \cite{Hetherington87} and
{\em (ii)} 
to extend our approach to negative values of $\lambda$ (following in that ref.~\cite{Pablo03})
increasing artificially the --- then not necessarily Gaussian --- fluctuations of the extensive variable and making 
thus the system increasingly unstable. 
%
%

\begin{acknowledgments}
H.X. thanks the CNRS and the IRTG Soft Matter for supporting her sabbatical stay in Strasbourg,
P.P., C.G. and J.H. the IRTG Soft Matter and F.W. the DAAD for funding.
We are indebted to A. Blumen (Freiburg) for helpful discussions.
\end{acknowledgments}

\appendix
\section{Affine interaction energy}
\label{app_affineenergy}

\paragraph*{Affine displacement assumption.}
We consider here the change of the conservative interaction energy $\Uexs(\Vhat)$ of a configuration $s$
under an imposed dilatational strain {\em assuming} that all particles respond to the macroscopic 
constraint by an {\em affine} microscopic displacement. 
%
%
It is convenient to introduce a dimensionless scalar characterizing the relative volume change 
\begin{equation}
\epsilon \equiv \delta \Vhat(\epsilon)/\Vhat(0) \equiv \Vhat(\epsilon)/\Vhat(0) -1
\label{eq_epsilondef}
\end{equation}
with $\Vhat(0)$ being the instantaneous volume of the unperturbed reference simulation box ($\epsilon=0$).
Obviously, for canonical ensembles the instantaneous volume $\Vhat$ can be replaced by $V$.
The affinity assumption implies, e.g.,
\begin{equation}
x(0) \Rightarrow x(\epsilon) = x(0) (1+\epsilon)^{1/d}
\label{eq_strain_x_epsilon}
\end{equation}
for the $x$-coordinate of particle positions or relative particle distances 
where the argument $(0)$ denotes again the reference system.
Equation~(\ref{eq_strain_x_epsilon}) implies that the squared distance $r^2$ between two particles transforms as
\begin{equation}
r^2(0) \Rightarrow r^2(\epsilon) = r^2(0) (1+\epsilon)^{2/d}.
\label{eq_strain_r_epsilon}
\end{equation}
It follows that
\begin{eqnarray}
\frac{d r^2(\epsilon)}{d\epsilon} & \to & \frac{2}{d} r^2(0), \label{eq_strain_rone_epsilon} \\
\frac{d^2 r^2(\epsilon)}{d\epsilon^2} & \to & \frac{2(2-d)}{d^2} r^2(0) \label{eq_strain_rtwo_epsilon}
\end{eqnarray}
where we have taken finally for both derivatives the limit $\epsilon \to 0$.
For the first two derivatives of a general function $f(r(\epsilon))$ with respect to $\epsilon$ this implies
\begin{eqnarray}
\frac{\partial f(r(\epsilon))}{\partial \epsilon} &  \to & \frac{1}{d} r f^{\prime}(r) \label{eq_strain_froneBeps} \\
\frac{\partial^2 f(r(\epsilon))}{\partial \epsilon^2} &  \to  &
\frac{1}{d^2} \left(r^2 f^{\prime\prime}(r) + r f^{\prime}(r) \right) \nonumber \\
 & - & \frac{1}{d} r f^{\prime}(r) \label{eq_strain_frtwoBeps}
\end{eqnarray}
for $\epsilon \to 0$ and dropping the argument $(0)$ on the right hand-sides.

\paragraph*{Pair potential assumption.}
Let us assume now in addition that the interaction energy is given by 
\begin{equation}
\Uexs(\epsilon) = \sum_l u(\rl(\epsilon))
\label{eq_Upairinteraction}
\end{equation}
where the index $l$ labels the interactions between the particles $i$ and $j$ with $i < j$.
We note {\em en passant} that for the first derivative of the interaction energy $\Uexs$ 
with respect to the volume eq.~(\ref{eq_strain_froneBeps}) implies
\begin{equation}
- \Uexs^{\prime}(\Vhat) = - \Uexs^{\prime}(\epsilon)/V(0) = 
- \frac{1}{d \Vhat} \sum_l \rl u^{\prime}(\rl)
\label{eq_UexsPex}
\end{equation}
where we have taken $\epsilon \to 0$ in the last step. As stated in the main text, 
eq.~(\ref{eq_PexKirkwood}), this is exactly the Kirkwood virial for the 
instantaneous excess pressure $\Pexhat$. 
As defined in eq.~(\ref{eq_etaAex}), the coefficient $\etaAex/V$ measures the second derivative 
of the interaction energy $\Uexs(\Vhat)$ of a configuration $s$ with respect to an affine 
dilatational strain. Using eq.(\ref{eq_strain_frtwoBeps}) it follows for the instananeous 
value $\etaAexhat$ that
\begin{eqnarray}
\etaAexhat/V & \equiv & \Uexs^{\prime\prime}(\Vhat) = \frac{1}{\Vhat(0)^2} \ \Uexs^{\prime\prime}(\epsilon) \nonumber \\ 
 & = &
\frac{1}{d^2\Vhat^2} \sum_l \rl^2 u^{\prime\prime}(\rl) + \rl u^{\prime}(\rl) \nonumber \\
&- & \frac{1}{d \Vhat^2} \sum_l \rl u^{\prime}(\rl) \label{eq_etaAexhat}
\end{eqnarray}
where we have taken again $\epsilon \to 0$ and have dropped the index $(0)$ in the last step.

\paragraph*{Simple averages.}
Only the affinity assumption and the pair potential choice have been used up to now.
The mean value $\etaAex = \langle \etaAexhat \rangle$ is then obtained by taking the thermal average 
over all configurations $s$ and over all volumes $\Vhat$ depending on the ensemble. 
Since all contributions to eq.~(\ref{eq_etaAexhat}) correspond to {\em simple averages},
thermostatistics tells us that one can replace $\Vhat$ for sufficiently large systems by 
its mean value $V$, eq.~(\ref{eq_Vhatn}). 
%
%
One confirms that the affine dilatational elasticity becomes
\begin{eqnarray}
\etaAex & = & \etaB + \Pex \mbox{ with } \label{app_etaAex} \\
\etaB & = & \frac{1}{d^2 V} \la \sum_l \rl^2 u^{\prime\prime}(\rl) + \underline{\rl u^{\prime}(\rl)} \ra 
\label{app_etaB}
\end{eqnarray}
as already stated in eq.~(\ref{eq_etaB}).
Note that it is inconsistent to neglect the explicit excess pressure contribution to $\etaAex$ in
eq.~(\ref{app_etaAex}) but to keep the underlined contribution to $\etaB$ which amounts to $-\Pex/d$.
The sum of both terms $\Pex (d-1)/d$ only vanishes in $d=1$ dimensions
where
\begin{equation}
\etaAex = \frac{1}{V} \la \sum_l \xl^2 u^{\prime\prime}(\xl) \ra
\label{eq_etaAexdimone}
\end{equation}
as used in sect.~\ref{simu_affine}.

\section{Volume rescaling trick revisited}
\label{app_volumerescal}

\paragraph*{Introduction.}
The stress fluctuation formula, eq.~(\ref{eq_Rowl}), for the compression modulus $K$
of isotropic systems has been derived in sec.~\ref{theo_Krowl} essentially using the 
general transformation relation eq.~(\ref{eq_dAdB}) between conjugated ensembles and 
evaluating the pressure fluctuations in the \NPT-ensemble by integration by parts, 
eq.~(\ref{eq_etaA}).
Historically, eq.~(\ref{eq_Rowl}) has been first derived properly by Rowlinson \cite{RowlinsonBook} 
who computed using the volume rescaling trick the second derivative, eq.~(\ref{eq_Kdef}), 
of the free energy $F(T,V)=-\kBT \log[Z(V)]$. 
(The \NVT-ensemble is thus used and, hence, $V=\Vhat$ in the following.)
Since the total system Hamiltonian may be written as the sum of a kinetic energy
and a potential energy, the partition function $Z(V)$ factorizes in an ideal contribution
$\Zid(V)$ and an excess contribution $\Zex(V)$ on which we focus below.
We remind \cite{Callen} that using $\Zid(V) \sim V^N$ one readily confirms 
$\Pid = \kBT \rho$ and $\Kid = \Pid$ for the ideal contributions to, respectively, 
the total pressure $P=\Pid+\Pex$ and the total compression modulus $K=\Kid+\Kex$.

\paragraph*{Mapping of strained and unstrained configurations.}
Since both the perturbed as the unpertubed partition function is sum over all possible particle
configurations, one can always compare the contribution of a configuration $s$
of the strained system with the contribution of a configuration of the reference 
being obtained by the affine rescaling of all coordinates according to eq.~(\ref{eq_strain_x_epsilon}).
Using the same notations as in appendix~\ref{app_affineenergy} the interaction energy $\Uexs(\epsilon)$ 
of the strained system can be expressed in terms of the coordinates (state) of the unperturbed system 
and the explicit metric parameter $\epsilon$. The excess contribution $\Fex(\epsilon) = -\kBT \Zex(\epsilon)$
is thus obtained from 
$\Zex(\epsilon) = \sum_s \exp\left[ -\beta \Uexs(\epsilon) \right]$
where constant prefactors have been omitted and where the sum is taken over all possible configurations $s$.

\paragraph*{General conservative potential.}
We note for the first two derivatives of the free energy 
\begin{eqnarray}
\left[ \ln(\Zex(\epsilon)) \right]^{\prime} & = & \frac{\Zex^{\prime}(\epsilon)}{\Zex(\epsilon)} \label{eq_dlogZdeps} \\
\left[ \ln(\Zex(\epsilon)) \right]^{\prime\prime} & = & \frac{\Zex^{\prime\prime}(\epsilon)}{\Zex(\epsilon)}
-\left(\frac{\Zex^{\prime}(\epsilon)}{\Zex(\epsilon)}\right)^2 \label{eq_ddlogZddeps}
\end{eqnarray}
and of the partition function
\begin{eqnarray}
\Zex^{\prime}(\epsilon) & = & - \sum_s \beta \Uexs^{\prime}(\epsilon) \ e^{-\beta \Uexs(\epsilon)} \label{eq_dZdeps} \\
\Zex^{\prime\prime}(\epsilon) & =
& \sum_s \left( \beta \Uexs^{\prime}(\epsilon) \right)^2 \ e^{-\beta \Uexs(\epsilon)} \nonumber \\
& - & \sum_s \left( \beta \Uexs^{\prime\prime}(\epsilon) \right) e^{-\beta \Uexs(\epsilon)}
\label{eq_ddZddeps}
\end{eqnarray}
with a prime denoting again a derivative with respect to the indicated argument.
Using $\Pex = -\partial \Fex(V)/\partial V$ and eq.~(\ref{eq_dZdeps}) and taking the limit 
$\epsilon \to 0$ one verifies that
\begin{equation}
\Pex = \la \Pexhat \ra \mbox{ with } \Pexhat \equiv -\frac{1}{V} \left.\Uexs^{\prime}(\epsilon)\right|_{\epsilon=0}
\label{eq_Pexhat_us}
\end{equation}
which {\em defines} the instantaneous excess pressure. 
(The average taken 
uses the weights of the unperturbed system.) The excess pressure thus measures the average change
of the total interaction energy $\Uexs(\epsilon)$ taken at $\epsilon=0$.
The excess compression modulus $\Kex$ is obtained using in addition eq.~(\ref{eq_ddlogZddeps}) and eq.~(\ref{eq_ddZddeps})
and taking finally the  $\epsilon \to 0$ limit. This yields
\begin{eqnarray}
\beta V \Kex & = & 
\left.\la \beta \Uexs^{\prime\prime}(\epsilon)\ra\right|_{\epsilon=0} \nonumber \\
& - & 
\left.\left(\la (\beta \Uexs^{\prime}(\epsilon))^2 \ra - \la \beta \Uexs^{\prime}(\epsilon) \ra^2  \right)\right|_{\epsilon=0}.  
\label{eq_KexUexs}
\end{eqnarray}
Being a simple average the first term on the right hand-side corresponds to $\etaAex$ as defined in eq.~(\ref{eq_etaAex}).
Using eq.~(\ref{eq_Pexhat_us}) the second term is seen to yield the pressure fluctuation contribution $\etaFexV$.
Summarizing all terms we have thus shown that
\begin{equation}
K = \Kid + \Kex = \Pid + \etaAex - \etaFexV
\label{eq_Rowlrestated}
\end{equation}
in agreement with eq.~(\ref{eq_Rowl}) if the Born-Lam\'e coefficient is defined more generally as 
$\etaB \equiv \etaAex - \Pex$.

\paragraph*{Pair potential choice.}
Up to now we have stated $\Pex$ and $\Kex$ for a general interaction potentiel $\Uexs(\epsilon)$.
Assuming the interactions to be described by a pairwise additive potential, eq.~(\ref{eq_Upairinteraction}),
the Kirkwood relation, eq.~(\ref{eq_PexKirkwood}), is confirmed using 
eq.~(\ref{eq_Pexhat_us}) and eq.~(\ref{eq_UexsPex}). As already noted at the end of appendix~\ref{app_affineenergy},
one confirms using eq.~(\ref{eq_etaAexhat}) that $\etaB = \etaAex - \Pex$ agrees with eq.~(\ref{eq_etaB}).

\section{General stress fluctuation formalism}
\label{app_generalmodul}

\paragraph*{Introduction.}
We remind here the general stress fluctuation formalism derived
by Squire, Hold and Hoover \cite{Hoover69} and show that Rowlinson's formula,
eq.~(\ref{eq_Rowl}), is a special case obtained by symmetry considerations.
For convenience we introduce the two linear projection operators
\begin{eqnarray}
\Ttwo[\Aab] & \equiv &
\frac{1}{d} \sum_{\alpha,\beta=1}^{d} \Aab \delta_{\alpha\beta} 
\label{eq_Ttwodef} \\
\Tfour[\Aabcd] & \equiv &
\frac{1}{d^2} \sum_{\alpha,\beta,\gamma,\delta=1}^{d} \Aabcd \delta_{\alpha\beta} \delta_{\gamma\delta}
\label{eq_Tfourdef}
\end{eqnarray}
with $\Aab$ and $\Aabcd$ being, respectively, second- and forth-rang tensors and 
$\delta_{\alpha\beta}$ the Kronecker symbol \cite{abramowitz}. 
Greek letters are used for the spatial coordinates $\alpha, \beta, \gamma, \delta = 1,\ldots,d$.
The following identities are readily verified
\begin{eqnarray}
\Tfour[\Aab B^{\gamma\delta}] & = & \Ttwo[\Aab] \ \Ttwo[B^{\alpha\beta}] \label{eq_Tfour_D} \\
\Tfour[\Aabcd] &= & 1 \mbox{ for } \Aabcd = \delta_{\alpha\beta} \delta_{\gamma\delta} \label{eq_Tfour_A} \\
\Tfour[\Aabcd] &= & 1/d \mbox{ for } \Aabcd = \delta_{\alpha\gamma} \delta_{\beta\delta} \label{eq_Tfour_B} \\
\Tfour[\Aabcd] &= & 1/d^2 \mbox{ for } \Aabcd = \na \nb \nc \nd \label{eq_Tfour_C}  
\end{eqnarray}
with $\na, \ldots \ $ being the spatial components of a normalized vector, i.e. $\sum_{\alpha} (\na)^2 = 1$.

\paragraph*{Thermodynamics and symmetry.}
%
Generalizing the definition of the pressure $P$ given in eq.~(\ref{eq_Kdef}),
the stress tensor $\sigmaTen$ may be defined as the first derivative of the free energy per volume
with respect to the linear strain $\epsilonTen$ \cite{LandauElasticity}
characterizing the macroscopic deformation of the system.
Note that in general $\sigmaTen$ does not vanish at the investigated state point, 
i.e. the systems may be {\em prestressed} at the reference strain $\epsilonTen=0$. 
The latter point is crucial for essentially all soft matter systems which only assemble 
because a finite density and/or stress is applied. It is less important for the classical
crystalline solids where the elastic moduli are normally huge compared to the imposed stresses.
We denote by $\delta \sigmaTen$ the increment of the stress tensor $\sigmaTen$ 
under an increment $\delta \epsilonTen$ 
generalizing the dilatational strain $\epsilon$ used in appendix~\ref{app_volumerescal}.
Assuming an infinitessimal strain increment, Hooke's law reads quite generally \cite{LandauElasticity}
\begin{equation}
\delta \sigmaTen = \sum_{\gamma,\delta=1}^{d} \Eabcd \delta \epsilon^{\gamma\delta}
\label{eq_Hooke}
\end{equation}
where the elastic moduli $\Eabcd$ stand for the second derivative 
$\partial^2 F/\partial \epsilon^{\alpha\beta} \partial \epsilon^{\gamma\delta}$
of the free energy per volume at the given thermodynamic state \cite{LandauElasticity}.
Let us now assume a pure dilatational strain without shear, i.e. $\delta \epsilonTen = \epsilon \ \delta_{\alpha\beta}$.
As may be seen from eq.~(4.6) of ref.~\cite{LandauElasticity}, this implies
$\delta \sigmaTen = d K \epsilon \delta_{\alpha\beta}$. Hence, $\sum_{\alpha} \delta \sigma^{\alpha\alpha} = d^2 K  \epsilon$. 
Or, using eq.~(\ref{eq_Hooke}) one sees that $\sum_{\alpha} \delta \sigma^{\alpha\alpha} = \sum_{\alpha \beta} E^{\alpha\alpha\beta\beta} \epsilon$.
Comparing both expressions, this shows that the compression modulus is given by $K = \Tfour[\Eabcd]$ 
using the linear projection operator defined above.

\paragraph*{Stress fluctuation relations.}
As described in the literature \cite{Hoover69,Lutsko89,FrenkelSmitBook,Barrat06,SBM11}, 
the elasticity tensor $\Eabcd$ can numerically be computed from the sum 
\begin{equation}
\Eabcd = \Babcd + \CKabcd + \CBabcd + \CFidabcd + \CFexabcd
\label{eq_Eabcd_sum}
\end{equation}
with contributions as specified below. 
The compression modulus is obtained by applying the linear operator $\Tfour$ to each term and 
by summing up the contributions.
Note that some authors only refer to the sum $\CKabcd + \CBabcd + \CFidabcd + \CFexabcd$
as the elasticity tensor \cite{Hoover69,Lutsko89,FrenkelSmitBook}.

\paragraph*{Initial stress contribution.}
The (often not included) first term $\Babcd$ in eq.~(\ref{eq_Eabcd_sum}) 
may we written as \cite{FrenkelSmitBook} 
\begin{eqnarray}
\Babcd & \equiv &  
-\sigma^{\alpha\beta} \delta_{\gamma\delta}
+ \sigma^{\alpha\gamma} \delta_{\beta\delta}
+ \sigma^{\alpha\delta} \delta_{\beta\gamma}
\label{eq_Babcd_gen} \\ 
& = & 
P \left(\delta_{\alpha\beta} \delta_{\gamma\delta} - \delta_{\alpha\gamma} \delta_{\beta\delta} - \delta_{\alpha\delta} \delta_{\beta\gamma} \right).
\label{eq_Babcd}
\end{eqnarray}
Following  Birch \cite{Birch37} we have assumed in the second step that the system is isotropically stressed,
i.e. $\sigma^{\alpha\beta} = - P \delta_{\alpha\beta}$. 
Since $\Babcd$ becomes negligible for small total pressures $P \approx 0$, this term is often not computed.
Consistency implies then, however, to set $\Pid=-\Pex$ in the remaining terms of eq.~(\ref{eq_Eabcd_sum}).
Returning to eq.~(\ref{eq_Babcd}) one sees using eq.~(\ref{eq_Tfour_A}) and eq.~(\ref{eq_Tfour_B}) that 
\begin{equation}
\Tfour[\Babcd] = P \left( 1 - \frac{2}{d}\right) = P - \frac{2 \Pid}{d}  \underline{- \frac{2 \Pex}{d}}.
\label{eq_BabcdTrace}
\end{equation}
There is thus no contribution to $K$ from the Birch term for $d=2$ dimensions and a contribution $-P$ for $d=1$.
The leading term $P$ in the second step of eq.~(\ref{eq_BabcdTrace}) corresponds to the total pressure 
indicated in eq.~(\ref{eq_Rowl}). 

\paragraph*{Ideal pressure contributions.}
The second and the forth term in eq.~(\ref{eq_Eabcd_sum}) correspond to the kinetic (ideal) contributions 
\begin{eqnarray}
\CKabcd & \equiv & 2 \Pid \left(\delta_{\alpha\gamma} \delta_{\beta\delta} + \delta_{\alpha\delta} \delta_{\beta\gamma} \right),
\label{eq_CKabcd} \\
\CFidabcd & \equiv & - \Pid \left(\delta_{\alpha\gamma} \delta_{\beta\delta} + \delta_{\alpha\delta} \delta_{\beta\gamma} \right). 
\label{eq_CFidabcd}
\end{eqnarray}
Using again eq.~(\ref{eq_Tfour_A}) and eq.~(\ref{eq_Tfour_B}) this yields $\Tfour[\CKabcd] = 4 \Pid/d$
and $\Tfour[\CFidabcd] = -2 \Pid/d$. Together with the ideal pressure contribution from eq.~(\ref{eq_BabcdTrace})
all the kinetic terms sum up (independently of the dimension) to $\Pid$ in agreement with eq.~(\ref{eq_Rowl}).
It is worthwhile to stress that if the Birch term $\Babcd$ is neglected for $\Pid \ne -\Pex$, 
the ideal pressure contributions sum up incorrectly as is the case in ref.~\cite{Hoover69}.
In fact, the traditional separation of the (rather trivial) different kinetic terms stated in eq.~(\ref{eq_Babcd}) 
is from the numerical point of view unfortunate. More importantly, this separation is theoretically misleading: 
The total system Hamiltonian being the sum of the kinetic energy and the conservative interaction potential, 
all kinetic contributions to the elastic moduli can (and should) be factorized out from the start.
This leads immediately to the obvious total kinetic contribution 
\begin{equation}
\Cidabcd = \Pid \delta_{\alpha\beta} \delta_{\gamma\beta} 
\label{eq_Cidabcd}
\end{equation}
replacing the aforementioned three terms.

\paragraph*{Affine Born contribution.}
Generalizing the Born coefficient $\etaB$ in eq.~(\ref{eq_Rowl}), the third term $\CBabcd$ in eq.~(\ref{eq_Eabcd_sum}) 
corresponds to the (second derivative) of the energy change assuming the affine displacement of all 
particles (as in appendix~\ref{app_affineenergy} for a pure dilatational strain). Assuming pair interactions
it is given by \cite{foot_consistKirkwood}
\begin{equation}
\CBabcd \equiv \frac{1}{V} \sum_{l} \la 
\left(\rl^2 u^{\prime\prime}(\rl) - \rl u^{\prime}(\rl) \right) \nla \nlb \nlc \nld \ra 
\label{eq_CBabcd}
\end{equation}
where $\nla$, $\ldots$ stand for the normalized vector between the pair of interacting particles
labeled by $l$.
Using eq.~(\ref{eq_Tfour_C}) one confirms that 
\begin{eqnarray}
\Tfour[\CBabcd] & = & \frac{1}{d^2V} \la \sum_l \rl^2 u^{\prime\prime}(\rl) - \rl u^{\prime}(\rl) \ra \nonumber \\
& = & \etaB + \underline{\frac{2\Pex}{d}}
\label{eq_CBabcdTrace}
\end{eqnarray}
where we have used the Kirkwood formula, eq.~(\ref{eq_PexKirkwood}), in the last step.
Note that the underlined term cancels exactly the underlined last term in eq.~(\ref{eq_BabcdTrace}).
This is equivalent to the statement made after eq.~(\ref{app_etaB}) in appendix~\ref{app_affineenergy}.

\paragraph*{Stress fluctuations.}
The last two terms indicated in eq.~(\ref{eq_Eabcd_sum}) stand for the
correlation function 
\begin{equation}
\CFabcd \equiv \CFidabcd + \CFexabcd 
\equiv -\beta V\la \delta \hat{\sigma}^{\alpha \beta} \delta \hat{\sigma}^{\gamma \delta} \ra
\label{eq_CFabcd}
\end{equation}
with $\delta \hat{\sigma}^{\alpha \beta}$ denoting the fluctuation of the total stress tensor 
$\sigmaTen = \sigmaidTen + \sigmaexTen$. 
Using the factorization of ideal and excess stress fluctuations in the canonical ensemble at constant volume $V$
(as shown in sect.~\ref{theo_NPT} this is different in \NPT-ensembles), both contributions 
can be separated. The ideal stress fluctuation contribution has already been accounted for above.
We remind that the instantaneous excess pressure $\Pexhat$ is given by the trace $\Pexhat = \Ttwo[\pressexTen]$ 
over the excess pressure tensor $\pressexTen$ being the negative of the instantaneous excess stress tensor. 
Using eq.~(\ref{eq_Tfour_D}) one thus obtains 
\begin{equation}
\Tfour[\CFexabcd] = - \beta V \la \Ttwo[\delta \pressexTen]^2 \ra
\label{eq_CFexabcdTrace}
\end{equation}
which is identical to the pressure fluctuation term in eq.~(\ref{eq_Rowl}).
We have thus confirmed that the general stress fluctuation relation, eq.~(\ref{eq_Eabcd_sum}), 
is consistent with the Rowlinson formula for the compression modulus $K$. 


\begin{thebibliography}{47}
\bibitem[{foo({\natexlab{b}})}]{foot_trunc}
\bibinfo{note}{If a truncated potential is used as for the two
  glass-forming liquids discussed, some care is needed for the computation of
  $\etaB$. 
  Being a moment of the second potential derivative,
  $\etaB$ needs to be corrected using a weighted histogram evaluated at the
  cutoff as described in ref.~\cite{XWP12}.  This correction is a
  simple average and the same value is obtained for any $\lambda$. 
  It becomes relevant if eq.~(\ref{eq_KrowlL}) is probed for small $\lambda$.}
\bibitem[{foo({\natexlab{c}})}]{foot_classical}
\bibinfo{note}{We only consider classical systems. The generalization to
  quantum systems may involve delicate problems.}
\bibitem[{foo({\natexlab{d}})}]{foot_critical}
\bibinfo{note}{The general stress fluctuation formalism may be applied close to
  a first order phase transition only as long as the transformation relation
  between conjugated ensembles \cite{Lebowitz67} remains valid. Necessary
  conditions are that the elastic modulus of interest remains positive definite
  and that the fluctuations of the extensive variable $X$ are symmetric around
  the main maximum of the distribution $\plam(\Xhat)$. The stress fluctuation
  formalism becomes incorrect in general close to a second order phase
  transition.}
\bibitem[{foo({\natexlab{e}})}]{foot_kextdim}
\bibinfo{note}{The spring constant $\kext$ has the dimension energy per
  volume$^2$ which implies, as one expects, $[\Kext] =$ energy/volume for the
  associated compression modulus.}
\bibitem[{foo({\natexlab{h}})}]{foot_intensive}
\bibinfo{note}{With $X$ being the extensive variable, the intensive variable
  may be either defined as the derivative $I = \partial U(X)/\partial X$ of the
  inner energy $U$ or as the derivative $J = \partial
  S(X)/\partial X$ of the entropy $S$ \cite{Callen}. It is the second
  definition which is used in ref.~\cite{Lebowitz67}. Note that $J = - \beta I$
  for all extensive variables $X$ other than $U$ \cite{Callen}. In
  our case we have $X=V$, $I= - P$ and $J= \beta P$.}
\bibitem[{foo({\natexlab{i}})}]{foot_ergodicity}
\bibinfo{note}{Ergodicity problems are irrelevant since these systems have been
  sampled by means of MC simulation.}
\bibitem[{foo({\natexlab{j}})}]{foot_cusp}
\bibinfo{note}{Qualitatively, this is similar to the predicted \cite{ZT13} and
  numerically observed \cite{WXP13,Barrat88} cusp-like singularity $G \approx
  \sqrt{1-T/\Tglass}$ of the shear modulus $G$ in colloidal and polymer glasses
  at the glass transition temperature $\Tglass$ due to the increase of the
  non-affine displacements.}
\bibitem[{foo({\natexlab{k}})}]{foot_etaFexVscaling}
\bibinfo{note}{It is possible to collapse the data by
  plotting the ratio of $\etaFexV$ and the second term in
  eq.~(\ref{eq_etaFexnet}) as function of the ratio of $\Pid$ and a crossover
  ideal pressure $\Pid^*$.}
\end{thebibliography}

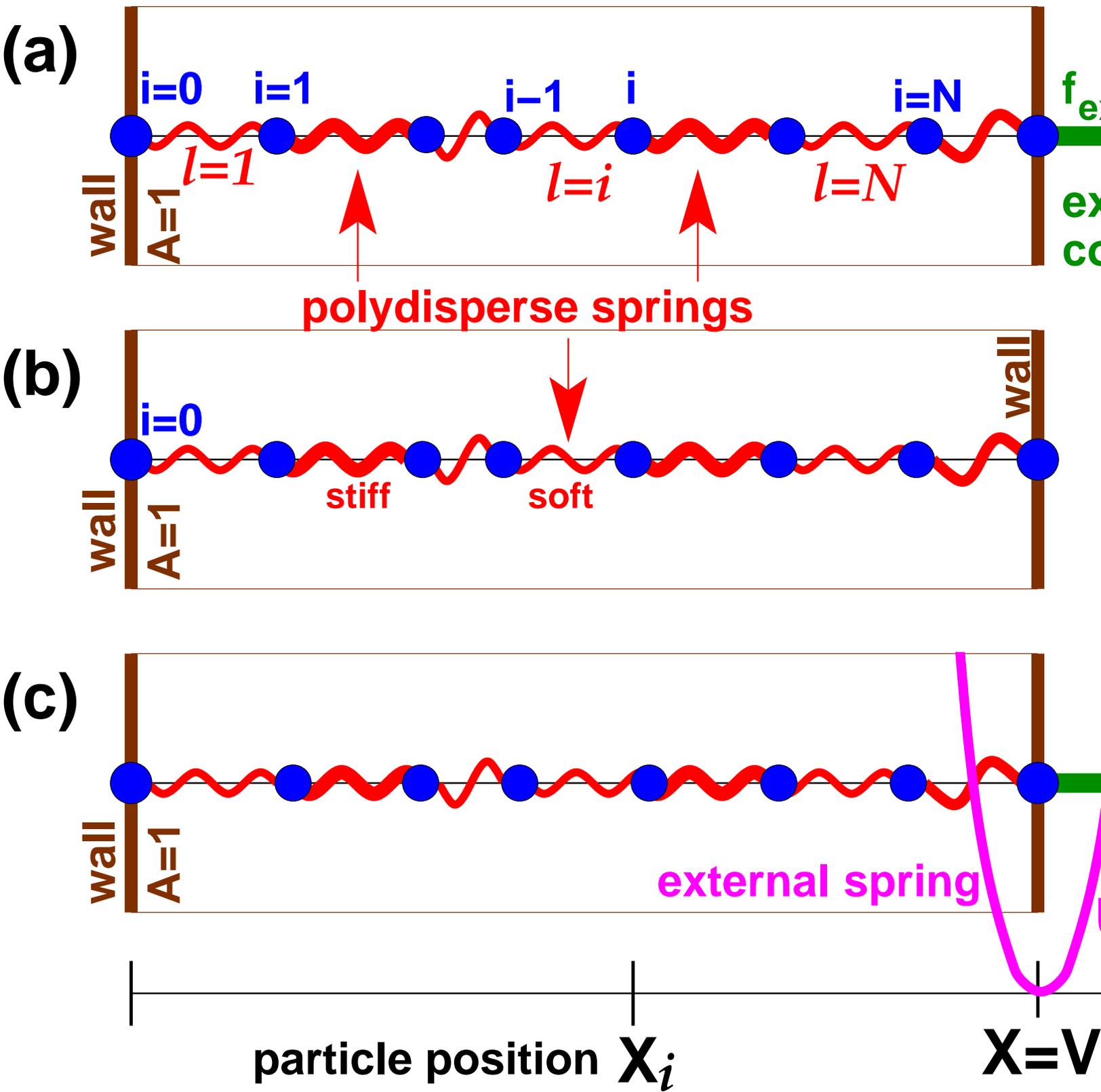
\end{document}